\documentstyle[arevdefs,epsf]{arevaa} 
\newcommand{\markcite}{\nocite}
\newcommand{\mch}{$M_{\rm chan}$~}

\begin{document}

\title{Type Ia Supernova Explosion Models} 
\markboth{Hillebrandt \& Niemeyer}{Type Ia Supernova Explosion Models} 
\author{Wolfgang Hillebrandt and Jens C. Niemeyer
\affiliation{Max-Planck-Institut f\"ur Astrophysik,
Karl-Schwarzschild-Str.~1, 85740 Garching, Germany}}  
 

\begin{abstract}

Because calibrated light curves of Type Ia supernovae
have become a major tool to determine the local expansion rate of the
Universe and also its geometrical structure, 
considerable attention has been given to models of these events
over the past couple of years. There are good reasons to believe
that perhaps most Type Ia supernovae are the explosions of white dwarfs that 
have approached the Chandrasekhar mass, \mch $\approx$ 1.39 M$_{\odot}$,
and are disrupted by thermonuclear fusion of carbon and oxygen.
However, the mechanism whereby such accreting carbon-oxygen white dwarfs
explode continues to be uncertain. 
Recent progress in modeling Type Ia supernovae as
well as several of the still open questions are addressed in this review.
Although the main emphasis will be
on studies of the explosion mechanism itself and on the related physical
processes, including the physics of turbulent nuclear combustion in
degenerate stars, we also discuss observational constraints.   
 
\end{abstract}

\maketitle

\section{INTRODUCTION}
  
Changes in the appearance of the night sky, visible with the naked eye,
have always called for explanations (and speculations).
But, although ``new stars'', i.e. novae and supernovae, are observed by humans   
for thousands of years, the modern era of supernova research began only 
about one century ago on August 31, 1885, when Hartwig discovered a ``nova''
near the center of the Andromeda galaxy, which became invisible about
18 months later. In 1919 Lundmark estimated the distance of M31 to be about
$7\times 10^5$ lyr, and by that time it became obvious that Hartwig's nova 
had been several 1000 times brighter than a normal nova \markcite{Lu20}({Lundmark} 1920).
It was also \markcite{Lu21}{Lundmark} (1921) who first suggested an association between
the supernova observed by Chinese astronomers in 1054 and the Crab nebula.
 
A similar event as S Andromeda was observed in 1895 in NGC 5253 (``nova''
Z Centauri), and this time the ``new star'' appeared to be 5 times 
brighter than the entire galaxy, but it was not before 1934 that a clear
distinction between classical novae and supernovae was made \markcite{BZ34}({Baade} \& {Zwicky} 1934).
Systematic searches, performed predominantly
by Zwicky, lead to the discovery of 54 supernovae in the years up to 1956 and,
due to improved observational techniques, 82 further supernovae were
discovered in the years from 1958 to 1963, all of course in external galaxies
\markcite{Zw65}(e.g., {Zwicky} 1965).
 
Until 1937 spectrograms of supernovae were very rare, and what was known seemed 
to be not too different from common novae. This changed with the very bright
(m$_V \simeq 8.4$) supernova SN1937c in IC 4182 which had spectral features
very different from any object that had been observed before \markcite{Po37}({Popper} 1937).
All of the
other supernovae discovered in the following years showed very little dispersion
in their maximum luminosity and their post-maximum spectra looked very
similar at a given time. Based on this finding \markcite{Wi39}{Wilson} (1939) and \markcite{Zw39}{Zwicky} (1938a) 
suggested to use supernovae as distance indicators.
 
In 1940 it became clear, however, that there exist at least two distinctly 
different classes of supernovae. SN1940c in NGC 4725 had a spectrum very different
from all other previously observed supernovae for which good data were available
at that time, leading \markcite{Mi40}{Minkowski} (1940) to introduce the names ``Type I'' for 
those with spectra like SN1937c and ``Type II'' for SN 1940c-like events,
representing supernovae without and with Balmer-lines of hydrogen 
near maximum light. 
 
Whether or no the spectral
differences also reflect a different explosion mechanism was not known.
In contrast, the scenario originally suggested by \markcite{Zw38}{Zwicky} (1938b),
that a supernova occurs as the transition from an ordinary star to
a neutron star and gains its energy from the gravitational binding of
the newly born compact object, was for many years the only explanation.
\markcite{HF60}{Hoyle} \& {Fowler} (1960) were the first to discover that thermonuclear
burning in an electron-degenerate stellar core might trigger an explosion
and (possibly) the disruption of the star. Together with the idea  
that the light curves could be powered by the decay-energy of
freshly produced radioactive $^{56}$Ni \markcite{TAC67,CM69}(Truran, Arnett, \& Cameron 1967; {Colgate} \& {McKee} 1969) this
scenario is now the generally accepted one for a sub-class of all
Type I supernovae called Type Ia today. It is a bit amusing to note
that all supernovae (besides the Crab nebula) on which Zwicky had
based his core-collapse hypothesis were in fact of Type Ia and most
likely belonged to the other group, whereas the first core-collapse
supernova, SN1940c, was observed only about one year after he published 
his paper.
 
To be more precise, supernovae which do not show
hydrogen lines in their spectra but a strong silicon P Cygni -- feature
near maximum light are named Type Ia \markcite{WH90}({Wheeler} \& {Harkness} 1990). They
are believed to be the result of thermonuclear disruptions of white
dwarfs, either consisting of carbon and oxygen with a mass close to
the Chandrasekhar-mass, or of a low-mass C+O core mantled by a layer
of helium, the so-called sub-Chandrasekhar-mass models (see the recent
reviews by \markcite{W97,WW94a,WW94b}{Woosley} (1997b); {Woosley} \& {Weaver} (1994a, 1994b) and \markcite{NYSe94,NINe97}{Nomoto} et~al. (1994b, 1997)). The
main arguments in favor of this 
interpretation include: (i) the apparent lack of neutron stars in some of
the historical galactic supernovae (e.g. SN1006, SN1572, SN1604);
(ii) the rather homogeneous appearance of this sub-class; (iii)
the excellent fits to the light curves, which can be obtained from the
simple assumption that a few tenths of a solar mass of $^{56}$Ni is
produced during the explosion; (iv) the  good agreement with 
the observed spectra of typical Type Ia supernovae. Several of these
observational aspects are discussed in some detail in Section 2,
together with their cosmological implications. Questions concerning 
the nature of the progenitor stars are addressed in Section 3, and
models of light curves and spectra are reviewed in Section 5.

But having good arguments in favor of a particular explosion scenario
does not mean that this scenario is indeed the right one. Besides that
one would like to understand the physics of the explosion, the fact
that the increasing amount of data also indicates that there is a
certain diversity among the Type Ia supernovae seems to contradict a
single class of progenitor stars or a single explosion
mechanism. Moreover, the desire for using them as distance indicator
makes it necessary to search for possible systematic deviations from
uniformity. Here, again, theory can make important contributions. In
Section 4, therefore, we discuss the physics of thermonuclear
combustion, its implementation into numerical models of exploding
white dwarfs, and the results of recent computer simulations. 
A summary and conclusions follow in Section 6.

\section{OBSERVATIONS}
\label{observations}

The efforts to systematically obtain observational data of SNe Ia near
and far have gained tremendous momentum in recent years. This is
primarily a result of the unequaled potential of SNe Ia to act as
``standardizable'' candles \markcite{BT92,RPK96,HPMe95,T98}({Branch} \& {Tammann} 1992; {Riess}, {Press}, \& {Kirshner} 1996; {Hamuy} et~al. 1995; {Tripp} 1998) for the
measurement of the cosmological expansion rate \markcite{HPSe96a,B98}({Hamuy} et~al. 1996b; {Branch} 1998) and
its variation with lookback time \markcite{PAGe99,SSPe98,RFCe98}({Perlmutter} et~al. 1999; {Schmidt} et~al. 1998; {Riess} et~al. 1998). For
theorists, this development presents  both a challenge to help
understand the correlations among the observables and an opportunity
to use the wealth of new data to constrain the zoo of existing
explosion models. There exist a number of excellent reviews about SNe
Ia observations in general \markcite{F97a}({Filippenko} 1997b), their spectral properties
\markcite{F97b}({Filippenko} 1997a), photometry in the IR and optical bands
\markcite{MCGe96,MBGe97}({Meikle} et~al. 1996; Meikle et~al. 1997), and their use for measuring the Hubble constant
\markcite{B98}({Branch} 1998). Recent books that cover a variety of observational and
theoretical aspects of type Ia supernovae are \markcite{RCI97}Ruiz-Lapuente, Canal, \& Isern (1997) and
\markcite{NT00}Niemeyer \& Truran (2000). Below, we highlight those aspects of SN Ia
observations that most directly influence theoretical model building
at the current time.
 
\subsection{General Properties}
\label{properties}

The classification of SNe Ia is based on spectroscopic features: the
absence of hydrogen absorption lines, distinguishing them from Type II
supernovae, and the presence of strong silicon lines in the early and
maximum spectrum, classifying them as Types Ia \markcite{WH90}({Wheeler} \& {Harkness} 1990).

The spectral properties, absolute magnitudes, and light curve shapes
of the majority of SN Ia are remarkably homogeneous, exhibiting only
subtle spectroscopic and photometric differences
\markcite{BT92,HPSe96b,B98}({Branch} \& {Tammann} 1992; {Hamuy} et~al. 1996c; {Branch} 1998).  It was believed until recently
that approximately 85\% of all observed events
belong to this class of normal \markcite{BFN93}(``Branch-normal'', {Branch}, {Fisher}, \& {Nugent} 1993)
SNe Ia, represented for example by SNe 1972E, 1981B, 1989B, and
1994D. However, the peculiarity rate can be as high
as 30 \% as suggested by \markcite{LFRe00}Li et~al. (2000). 

The optical spectra of normal SN Ia's contain neutral and singly
ionized  lines of Si, Ca, Mg, S, and O at maximum light, indicating
that the outer layers of the ejecta are mainly composed of
intermediate mass elements \markcite{F97a}({Filippenko} 1997b). Permitted Fe II lines
dominate the spectra roughly two weeks after maximum when the
photosphere  begins to penetrate Fe-rich ejecta \markcite{H91}(Harkness 1991).  In the
nebular phase of the light curve tail, beginning approximately one
month after peak brightness, forbidden Fe  II, Fe III, and Co III
emission lines become the dominant spectral features \markcite{A80}({Axelrod} 1980). Some
Ca II remains observable in absorption even at late times
\markcite{F97b}({Filippenko} 1997a). The decrease of Co lines \markcite{A80}({Axelrod} 1980) and the relative
intensity of Co III and Fe III \markcite{KKPe94}({Kuchner} et~al. 1994) give evidence that the
light curve tail is powered by radioactive decay of $^{56}$Co
\markcite{TAC67,CM69}(Truran et~al. 1967; {Colgate} \& {McKee} 1969).

The early spectra can be explained by resonant scattering of a thermal
continuum  with P Cygni-profiles whose absorption component is
blue-shifted according to ejecta velocities of up to a few times $10^4$
km/s, rapidly  decreasing with time in the early phase
\markcite{F97b}({Filippenko} 1997a). Different lines have different expansion velocities
\markcite{PBCe96}({Patat} et~al. 1996), suggesting a layered structure of the explosion
products.

Photometrically, SN Ia rise to maximum light in the period of
approximately 20 days \markcite{RFWe99b}({Riess} et~al. 1999b) reaching
\begin{equation}
M_{\rm B} \approx M_{\rm V} \approx -19.30 \pm 0.03 + 5 \log(H_0/60)
\end{equation}
with a dispersion of $\sigma_M \le 0.3$ \markcite{HPSe96a}({Hamuy} et~al. 1996b). It is
followed by a first rapid  decline of about three magnitudes in a
matter of one month. Later, the light curve tail falls off in an
exponential manner at a rate of approximately one magnitude per
month. In the I-band, normal SNe Ia rise to a second maximum
approximately two days after the first maximum \markcite{MBGe97}(Meikle et~al. 1997).

It is especially interesting that the two most abundant elements in
the universe, hydrogen and helium,  so far have not been unambiguously
detected in SN Ia spectra (\markcite{F97b}{Filippenko} (1997a), but see \markcite{MCGe96}{Meikle} et~al. (1996) for a
possible identification of He) and there are no indications for radio
emission of SNe Ia. \markcite{CLSe96}{Cumming} et~al. (1996) failed to find any signatures of H
in the   early-time spectrum of SN 1994D and used this fact to
constrain the mass accretion rate of of the progenitor wind
\markcite{LC97}({Lundqvist} \& {Cumming} 1997). The later spectrum of SN 1994D also did not exhibit
narrow H$\alpha$ features \markcite{F97a}({Filippenko} 1997b). Another direct constraint for
the progenitor system accretion rate comes from the non-detection of
radio emission from SN 1986G \markcite{ECRe95}({Eck} et~al. 1995), used by \markcite{BB95}{Boffi} \& {Branch} (1995) to
rule out symbiotic systems as a possible progenitor of this event.

\subsection{Diversity and Correlations}

Early suggestions \markcite{P77,B81}({Pskovskii} 1977; {Branch} 1981) that the existing inhomogeneities
among SN Ia observables are strongly intercorrelated are now
established beyond doubt \markcite{HPSe96c,F97b}({Hamuy} et~al. 1996a; {Filippenko} 1997a). \markcite{B98}{Branch} (1998) offers a
recent summary of correlations  between spectroscopic line strengths,
ejecta velocities, colors, peak absolute magnitudes, and light curve
shapes. Roughly speaking, SNe Ia appear to be arrangeable in a
one-parameter sequence according to explosion strength, wherein the
weaker explosions are less luminous, redder, and have a faster
declining light curve and slower ejecta velocities than the more
energetic events \markcite{B98}({Branch} 1998). The relation between the width of the
light curve around maximum and the peak brightness is the most
prominent of all correlations \markcite{P77,P93}({Pskovskii} 1977; {Phillips} 1993). Parameterized either by
the decline rate $\Delta m_{15}$ \markcite{P93,HPSe96c}({Phillips} 1993; {Hamuy} et~al. 1996a), a ``stretch
parameter'' \markcite{PGGe97}({Perlmutter} et~al. 1997), or a multi-parameter nonlinear fit in
multiple colors \markcite{RPK96}({Riess} et~al. 1996), it was used to renormalize the peak
magnitudes of a variety of observed events, substantially reducing the
dispersion of absolute brightnesses \markcite{RPK96,T98}({Riess} et~al. 1996; {Tripp} 1998). This correction
procedure is a central ingredient of all current cosmological surveys
that use SNe Ia as distance indicators \markcite{PAGe99,SSPe98}({Perlmutter} et~al. 1999; {Schmidt} et~al. 1998).

SN 1991bg and SN 1992K are well-studied examples for red, fast, and
subluminous supernovae \markcite{FRBe92,LKPe93,HPMe94,TBCe96}({Filippenko} et~al. 1992a; {Leibundgut} et~al. 1993; {Hamuy} et~al. 1994; {Turatto} et~al. 1996). Their V,
I, and R-band light  curve declined unusually quickly, skipping the
second maximum in I, and their spectrum showed a high  abundance of
intermediate mass elements (including Ti II) with low expansion
velocities but only little iron \markcite{FRBe92}({Filippenko} et~al. 1992a). Models for the nebular
spectra and light curve of SN 1991bg consistently imply that the total
mass of $^{56}$Ni in the ejecta was very low ($\sim 0.07$ M$_\odot$)
\markcite{MCTe97}({Mazzali} et~al. 1997a).  On the other side of the luminosity function, SN
1991T is typically mentioned as the most striking  representative of
bright, energetic events with broad light curves
\markcite{PWSe92,JLKe92,FRMe92,RCTe92,SMAe92}({Phillips} et~al. 1992; {Jeffery} et~al. 1992; {Filippenko} et~al. 1992b; {Ruiz-Lapuente} et~al. 1992; {Spyromilio} et~al. 1992). Rather than the expected
Si II and Ca II, its early spectrum displayed high-excitation lines of
Fe III but returned to normal a few months after maximum
\markcite{FRMe92}({Filippenko} et~al. 1992b).

Peculiar events like SN 1991T and SN 1991bg were suggested to belong
to different subgroups of SNe Ia than the normal majority, created by
different explosion mechanisms \markcite{MCTe97,FRMe92,FBHe99}({Mazzali} et~al. 1997a; {Filippenko} et~al. 1992b; {Fisher} et~al. 1999). The
overall SN Ia luminosity function seems to be very steep on the bright
end \markcite{VBMe95}({Vaughan} et~al. 1995), indicating that ``normal'' events are essentially
the brightest while the full class may contain a large number of
undetected subluminous SNe Ia \markcite{L99}(Livio 2000). New results \markcite{LFRe00}(Li et~al. 2000)
indicate, however, that the 
luminosity function may be shallower than anticipated.

There is also mounting evidence that SN Ia observables are correlated
with the host stellar population \markcite{B98}({Branch} 1998). SNe Ia in red or
early-type galaxies show, on average, slower ejecta velocities, faster
light curves, and are dimmer by $\approx 0.2$ to 0.3 mag than those in
blue or late-type galaxies \markcite{HPMe95,HPSe96c,BRB96}({Hamuy} et~al. 1995, 1996a; {Branch}, {Romanishin}, \& {Baron} 1996). The SN Ia
rate per unit luminosity is nearly a factor of two higher in late-type
galaxies than in early-type ones \markcite{CTTe97}({Cappellaro} et~al. 1997). In addition, the
outer regions of  spirals appear to give rise to similarly dim SNe Ia
as ellipticals whereas the inner regions harbor a wider variety of
explosion strengths \markcite{WHW97}({Wang}, {H\"oflich}, \& {Wheeler} 1997). When corrected for the difference
in light curve shape, the variation of absolute magnitudes with galaxy
type vanishes along with the dispersion of the former. This fact is
crucial for cosmological SN Ia surveys, making the variations with
stellar population consistent with the assumption of a single
explosion strength parameter \markcite{PAGe99,RFCe98}({Perlmutter} et~al. 1999; {Riess} et~al. 1998).

\subsection{Nearby and Distant SNe Ia}

Following a long and successful tradition of using relatively nearby
($z \le 0.1$,  comprised mostly of the sample discovered by the
Cal\'an/Tololo survey \markcite{HPSe96c}({Hamuy} et~al. 1996a)) SNe Ia for determining the
Hubble constant \markcite{B98}({Branch} 1998),  the field of SN Ia cosmology has
recently seen a lot of activity expanding the range of observed events
out to larger redshift, $z \approx 1$. Systematic searches involving a
series of wide-field images taken at epochs separated by 3-4 weeks, in
addition to pre-scheduled follow-up observations to obtain detailed
spectroscopy and photometry of selected events, have allowed two
independent groups of observers  --  the Supernova Cosmology Project
(SCP) \markcite{PAGe99}({Perlmutter} et~al. 1999) and the High-$z$  Supernova Search Team
\markcite{SSPe98}({Schmidt} et~al. 1998) -- to collect data of more than 50 high-redshift
SNe. Extending the Hubble diagram out to $z \approx 1$ one can, given
a sufficient number of data points over a wide range of $z$, determine
the density parameters for matter and cosmological constant,
$\Omega_{\rm M}$ and $\Omega_\Lambda$, independently \markcite{GP95}({Goobar} \& {Perlmutter} 1995) or,
in other words, constrain the equation of state of the universe
\markcite{GJCe98}({Garnavich} et~al. 1998). Both groups come to a spectacular conclusion
\markcite{RFCe98,PAGe99}({Riess} et~al. 1998; {Perlmutter} et~al. 1999): The distant SNe are too dim by $\approx 0.25$
mag to be consistent with a purely matter dominated, flat or open FRW
universe. Interpreted as being a consequence of a larger than expected
distance, this discrepancy can be resolved only if $\Omega_\Lambda$ is
non-zero, implying the existence of an energy component with negative
pressure. In fact, the SN Ia data is consistent with a spatially flat
universe made up of two parts vacuum energy and one part matter.

Both groups discuss in detail the precautions that were taken to avoid
systematic contaminations of the detection of cosmological
acceleration, including SN Ia evolution, extinction, and
demagnification by gravitational lensing. All of these effects would,
in all but the most contrived scenarios, give rise to an increasing
deviation from the $\Omega_\Lambda = 0$-case for higher redshift,
while the effect of a non-zero cosmological constant should become
less significant as $z$ grows. Thus, the degeneracy between a
systematic overestimation of the intrinsic SN Ia luminosity and
cosmological acceleration can be broken when sufficiently many events
at $z \ge 0.85$ are observed \markcite{FR99}({Filippenko} \& {Riess} 2000). Meanwhile, the only way to
support the cosmological interpretation is by ``\dots adding to the
list of ways in which they are similar while failing to discern any
way in which they are different'' \markcite{RFWe99a}({Riess} et~al. 1999a). This program has
been successful until recently: The list of similarities between
nearby and distant SNe Ia includes spectra near maximum brightness
\markcite{RFCe98}({Riess} et~al. 1998) and the distributions of brightness differences, light
curve correction factors, and $B-V$ color excesses of both samples
\markcite{PAGe99}({Perlmutter} et~al. 1999). Moreover, while the nearby sample covers a range of
stellar populations similar to the one expected out to $z \approx 1$,
a separation of the low-$z$ data into sub-samples  arising from
different progenitor populations shows no systematic shift of the
distance estimates \markcite{FR99}({Filippenko} \& {Riess} 2000). However, a recent comparison of the
rise times of more than 20 nearby SNe \markcite{RFWe99b}({Riess} et~al. 1999b)  with those
determined for the SCP high-redshift events gives preliminary evidence
for a difference of roughly 2.5 days.  This result was disputed by
\markcite{AKN00}{Aldering}, {Knop}, \& {Nugent} (2000) who conclude that the rise times of local and
distant supernovae are statistically consistent.

\subsection{Summary: Observational Requirements for Explosion Models}
\label{requ}

To summarize the main observational constraints, any viable scenario
for the SN Ia explosion mechanism has to satisfy the following
(necessary but probably not sufficient) requirements:
\begin{enumerate}
\item Agreement of the ejecta composition and velocity with observed
spectra and light curves. In general, the explosion must be
sufficiently powerful (i.e., produce enough $^{56}$Ni) and produce a
substantial amount of high-velocity intermediate mass elements in the
outer layers. Furthermore, the isotopic abundances of ``normal'' SNe
Ia must not deviate significantly from those found in the solar system.
\item Robustness of the explosion mechanism. In order to account for
the homogeneity of normal SNe Ia, the standard model should not give
rise to widely different outcomes depending on the fine-tuning of
model parameters or initial conditions.
\item Intrinsic variability. While the basic model should be robust
with respect to small fluctuations, it must contain at least one
parameter that can plausibly account for the observed sequence of
explosion strengths.
\item Correlation with progenitor system. The explosion strength
parameter must be causally connected with the state of the progenitor
white dwarf in order to explain the observed variations as a function
of the host stellar population.
\end{enumerate}

\section{LIGHT CURVE AND SPECTRA MODELING}
\label{radtrans}

Next we have to discuss the problem of coupling 
the interior physics of an exploding white dwarf to what is finally 
observed, namely light curves and spectra, by means of radiative
transfer calculations. 
For many astrophysical applications this problem is not solved, and
SN Ia are no exceptions. In fact, radiation transport is
even more complex in Type Ia's than for most other cases.   

A rough sketch of the processes involved can illustrate some of the  
difficulties (see, e.g., \markcite{ML93,EP93}{Mazzali} \& {Lucy} (1993); {Eastman} \& {Pinto} (1993)). Unlike most other   
objects we know in astrophysics SN Ia do not contain any
hydrogen. Therefore the opacities are always dominated either by
electron scattering (in the optical) or by a huge number of atomic 
lines (in the UV). In the beginning, the supernova is an opaque
expanding sphere of matter into which energy is injected from radioactive
decay. This could happen in a very inhomogeneous manner, as will      
discussed later. As the matter expands diffusion times eventually   
get shorter than the expansion time and the supernova becomes 
visual. However, because the star is rapidly                          
expanding the Doppler-shift of atomic lines causes important            
effects. For example, a photon emitted somewhere in the supernova may   
find the surrounding matter more or less transparent until it finds     
a line Doppler-shifted such that it is trapped in that line and       
scatters many times. As a consequence, the spectrum might look thermal
although the photon ``temperature'' has nothing in common with the    
matter temperature.                                                  
  
It is also obvious that radiation transport in SN Ia is very
non-local and that the methods used commonly in models of stellar     
atmospheres need refinements. As a consequence, there is no             
agreement yet among the groups modeling light-curves and spectra as to  
what the best approach is. Therefore it can   
happen that even if the same model for the interior physics of the    
supernova is inserted into one of the existing codes for modeling     
light-curves and spectra the predictions for what should be          
``observed'' could be different, again a very unpleasant situation.      
Things get even worse because all such models treat the exploding star
as being spherically symmetric, an assumption that is at least   
questionable, given the complex combustion physics discussed below.

In the following subsections we outline some of the commonly used
numerical techniques and also discuss their predictions for
SN Ia spectra and light curves. For more details on the techniques used by
the various groups, we refer readers to the articles by
\markcite{E97,Bl97,P97,BHM97,MTCe97,HKWe97}{Eastman} (1997); {Blinnikov} (1997); {Pinto} (1997); {Baron}, {Hauschildt}, \& {Mezzacappa} (1997); {Mazzali} et~al. (1997b); {H\"oflich} et~al. (1997), and \markcite{R97}{Ruiz-Lapuente} (1997). 
 
\subsection{Radiative transfer in Type Ia supernovae} 

In principle, the equations which have to be solved are well-known, 
either in form of the Boltzmann transport equation for photons or as a
transport equation for the monochromatic intensities. However, to
solve this time dependent, frequency dependent radiation transport
problem, including the need of treating the atoms in non-LTE, is very
expensive, even in spherically symmetric situations. Therefore,
approximations of various kinds are usually made which give rise
to controversial discussions. 
     
Conceptually, it is best to formulate and solve the transport equation 
in the co-moving (Lagrangian) frame (cf. \markcite{MM84}{Mihalas} \& {Weibel Mihalas} (1984)).
This makes the transport equation appear simpler, but causes problems
in calculating the ``co-moving'' opacity, in particular if the effect
of spectral lines on the opacity of an expanding shell of matter is
important, as in the case of SN Ia \markcite{KLCe77}({Karp} et~al. 1977).

There are different ways to construct approximate solutions of  
the transport equation. One can integrate over frequency and replace
the opacity terms by appropriate means, leaving a single (averaged)
transport equation. Unfortunately, in order to compute 
the flux-mean opacity one has to know the solution of 
the transport equation. Frequently the flux-mean is replaced,
for example, by the Rosseland mean, allowing for solutions,
but at the expense of consistency (see, e.g., \markcite{E97}{Eastman} (1997)).

Another way out is to replace the transport equation by its moment 
expansion introducing, however, the problem of closure. In its
simplest form, the diffusion approximation, the radiation field is
assumed to be isotropic, the time rate of change of the flux is
ignored, and the flux is expressed in terms of the gradient of the
mean intensity of the radiation field. Replacing the mean
intensity by the Planck function and closing the moment expansion by
relating the radiation energy density and pressure via an Eddington
factor (equal to 1/3 for isotropic radiation) finally leads to a set 
of equations that can be solved \markcite{MM84}({Mihalas} \& {Weibel Mihalas} 1984).

Again, this simple approach has several short-comings that are
obvious. First of all, the transition from an optically thick 
to thin medium at the photosphere requires a special treatment 
mainly because the radiation field is no longer isotropic. 
One can compensate for this effect by either putting in a 
flux-limiter or a variable Eddington factor to describe the 
transition from diffusion to free streaming, but both approaches 
are not fully satisfactory since it is difficult to calibrate 
the newly invented parameters (e.g., \markcite{Ku84,Fu87,BN91,MHHe92,
SMN92,YM95}{Kunasz} (1984); {Fu} (1987); {Blinnikov} \& {Nadyoshin} (1991); {Mair} et~al. (1992); {Stone}, {Mihalas}, \& {Norman} (1992); {Yin} \& {Miller} (1995)).

Alternatively, one can bin frequency space into groups and solve the
set of fully time dependent coupled monochromatic transport equations
for each bin. In this approach the problem remains to compute
average opacities for each frequency bin. Moreover, because of
computer limitations, in all practical applications the number of bins
cannot be large which introduces considerable errors, given the
strong frequency dependence of the line-opacities \markcite{BN91,E97}({Blinnikov} \& {Nadyoshin} 1991; {Eastman} 1997)
(see also Fig. 1).

Finally, in order to get synthetic spectra one might apply
Monte-Carlo techniques, as was done by \markcite{MTCe97}{Mazzali} et~al. (1997b) and \markcite{Lu99}{Lucy} (1999).
Here the assumption is that the supernova envelope is in homologous
spherical expansion and that the luminosity and the photospheric
radius are given. The formation of spectral lines is then computed
by considering the propagation of a wave packet emitted from the
photosphere subject to electron scattering and interaction with lines.
Line formation is assumed to occur by coherent scattering, and the
line profiles and escape probabilities are calculated in the 
Sobolev approximation. While this approach appears to be a powerful
tool to get synthetic spectra it lacks consistency since the
properties of the photosphere have to be calculated be other means.
 
\begin{figure*}
\epsfxsize=0.80\textwidth\epsfbox{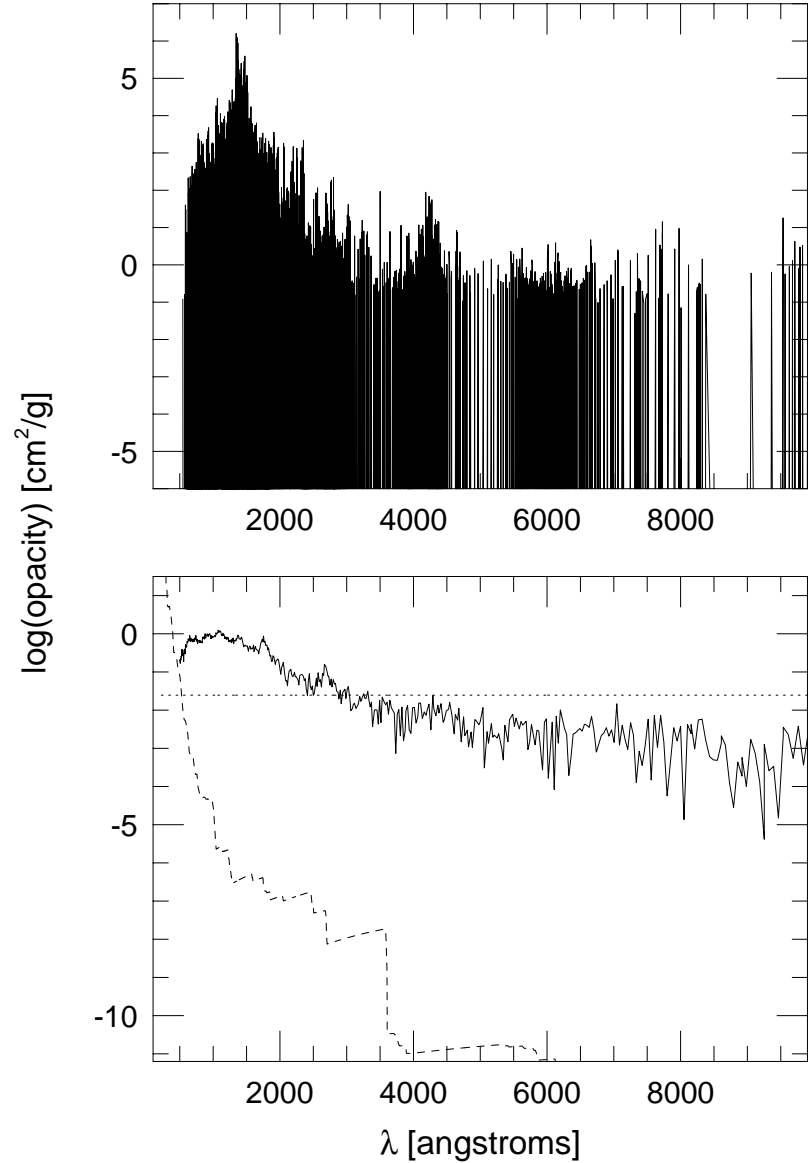}
\caption{\label{fig2} Mass opacity in a hot (23 000K) plasma of   
cobalt at a density of 10$^{-12}$g/cm$^3$. The upper plot shows
the line opacities (calculated by Iglesias, Rogers \& Wilson (1987)   
with OPAL), the lower one the bound-free (long dashed),
electron scattering (short dashed), and the  
``line expansion opacity'' (solid curve). (Courtesy of 
R. Eastman (1999), personal communication)} 
\end{figure*}

But having a numerical scheme at hand to solve the transport equation 
is not sufficient. It is even more important to have accurate
opacities. The basic problem, namely that at short wavelengths the
opacity is dominated by a huge number of weak lines, was mentioned
before. In practice this means that because the list included in
anyone's code is certainly incomplete and the available information may
not always be very accurate it is difficult to estimate possible
errors. Moreover, there is no general agreement among the different
groups calculating SN Ia lightcurves and spectra on how to correct
the opacities for Doppler-shifts of the lines, caused by the 
expansion of the supernova. The so-called ``expansion opacity'' 
(see  Fig. 1) that
should be used in approaches based on the diffusion equation as well
as on moment expansions of the transport equation is still discussed
in a controversial manner \markcite{PDMe96,Bl97,BHM97,E97,HKWe97,MTCe97,
P97}({Pauldrach} et~al. 1996; {Blinnikov} 1997; {Baron} et~al. 1997; {Eastman} 1997; {H\"oflich} et~al. 1997; {Mazzali} et~al. 1997b; {Pinto} 1997).

Other open questions include the relative importance of absorption and
scattering of photons in lines, and whether or not one can calculate
the occupation numbers of atomic levels in equilibrium (LTE) or has to
do it by means of the Saha-equation (NLTE) \markcite{PDMe96,NBBe97,HWT98}({Pauldrach} et~al. 1996; {Nugent} et~al. 1997; {H\"oflich}, {Wheeler}, \&  {Thielemann} 1998b).

\subsection{Results of numerical studies}

Despite of the problems discussed in the previous Subsection radiation
hydrodynamic models have been used widely as a diagnostic tool for
SN Ia. These studies include computations of $\gamma$-ray
\markcite{BT90,MHK91,BSR91,SKYe93,RLLe93,TW97,HWK98,WHLe99}({Burrows} \& {The} 1990; {M\"uller}, {H\"oflich}, \& {Khokhlov} 1991; {Burrows}, {Shankar}, \& {van Riper} 1991; {Shigeyama} et~al. 1993; {Ruiz-Lapuente} et~al. 1993b; {Timmes} \& {Woosley} 1997; {H\"oflich}, {Wheeler}, \&  {Khokhlov} 1998a; {Watanabe} et~al. 1999),
UV and optical \markcite{BV86,RCTe92,NBHE95,PDMe96,HKWe97,NBBe97,
HBB98,HWT98,BBHe99,FBHe99,Lu99,LBBe99}({Branch} \& {Venkatakrishna} 1986; {Ruiz-Lapuente} et~al. 1992; {Nugent} et~al. 1995; {Pauldrach} et~al. 1996; {H\"oflich} et~al. 1997; {Nugent} et~al. 1997; {Hatano}, {Branch}, \& {Baron} 1998; {H\"oflich} et~al. 1998b; {Lentz} et~al. 2000; {Fisher} et~al. 1999; {Lucy} 1999; {Lentz} et~al. 1999), 
and of infrared lightcurves and spectra \markcite{SPE94,Ho95,WHHe98}({Spyromilio}, {Pinto}, \& {Eastman} 1994; {H\"oflich} 1995; {Wheeler} et~al. 1998).
All studies are
based on the assumption, that the explosion remains on average
spherically symmetric, an assumption which is questionable, as
will be discussed in Sect. 5. Although spherical symmetry
might be a good approximation for temperatures, densities,
and velocities, the spatial distribution of the products
of explosive nuclear burning is expected to be very non-spherical,
and it is the distribution of the heavier elements, both in
real and in velocity space, which determines to a large extent
lightcurves and spectra.
 
With the possible exception of SN 1991T, where a 2 -- 3$\sigma$ detection
of the $^{56}$Co decay-lines at 847keV and 1238keV has been reported
\markcite{MBBe97}({Morris} et~al. 1997) (see, however, \markcite{LJKe95}{Leising} et~al. (1995)), only upper limits on 
 $\gamma$-ray line-emission from SN Ia are known. 
On the basis of the models this is not surprising
since the flux limits of detectors such as COMPTEL on GRO (~10$^{-5}$
photons per cm$^2$ and second \markcite{SBBe96}({Sch\"onfelder} et~al. 1996)) 
allows detections out to distances of about 15 Mpc
in the most favorable cases, i.e. delayed-detonation models
producing lots of $^{56}$Ni in the outer parts of the supernova
\markcite{TW97}({Timmes} \& {Woosley} 1997). In fact,
the tentative detection of decay-lines from SN 1991T
at a distance of about 13 Mpc can be explained by certain 
delayed-detonation models and was even predicted by some of
them (\markcite{MHK91}{M\"uller} et~al. (1991); see also Sect. 5).
 
Synthetic (optical and UV) spectra of hydrodynamic models of SN Ia 
have been computed by several groups \markcite{HK96,NBBe97}({H\"oflich} \& {Khokhlov} 1996; {Nugent} et~al. 1997) and have 
been compared with the observations. The bottom line of these
investigations is that Chandrasekhar-mass deflagration models
are in good agreement with observations of ``Branch-normals''
such as SN 1992A and SN 1994D \markcite{HK96,NBBe97}({H\"oflich} \& {Khokhlov} 1996; {Nugent} et~al. 1997), and delayed-detonation
are equally good. The reason is that in both classes of models 
the burning front starts by propagating out slowly, giving
the star some time to expand. The front then speeds up to higher
velocities, e.g. to a fair fraction of the sound velocity for
deflagration models and to supersonic velocity for detonations,
which is necessary to match the obseved high velocities 
of the ejecta. But as far as the amount
of radioactive Ni is concerned, the predictions 
of both classes of models are not
too different \markcite{NBBe97}({Nugent} et~al. 1997). It also appears
that sub-Chandrasekhar models cannot explain the observed UV-flux
and the colors of normal SNe Ia \markcite{KMH93}({Khokhlov}, {M\"uller}, \& {H\"oflich} 1993). Moreover,
although sub-Chandrasekhar models eject considerable amounts
of He, according to the synthetic spectra
He-lines should not be seen, eliminating them as a
tool to distinguish between the models \markcite{NBBe97}({Nugent} et~al. 1997).
  
In the infrared SN Ia do show non-monotonic behavior 
\markcite{EMNe85}({Elias} et~al. 1985) and, as for the bolometric lightcurves, 
a correlation between peak-brightness
and lightcurve-shape seems to exist \markcite{CL98,Co00}({Contardo} \& {Leibundgut} 1998; {Contardo} 1999). Therefore calculations
of IR lightcurves and spectra are of importance and they might
prove to be a good diagnostic tool. Broad-band IR lightcurves
have been computed by \markcite{HKW95a}{H\"oflich}, {Khokhlov}, \& {Wheeler} (1995) with the result that 
the second IR-peak can be explained as an opacity effect.
Although the fits were not perfect the general behavior, again,
was consistent with both, the deflagration and the delayed-detonation
models. Detailed early IR-spectra have been calculated only recently
\markcite{WHHe98}({Wheeler} et~al. 1998) and the models provide a good physical 
understanding of the spectra. Again, a comparison between  
several of the delayed-detonation models and SN 1994D gave good
agreement, but one might suspect that certain deflagrations
would do equally well. However, in principle, synthetic IR spectra 
are sensitive to the boundary between explosive C and O and 
between complete and incomplete Si burning \markcite{WHHe98}({Wheeler} et~al. 1998) and 
should provide some information on the progenitors and the 
explosion mechanism.
 
In conclusion, models of SN Ia lightcurves and spectra can
fit the observations well but, so far, their predictive
power is limited. The fact that multi-dimensional effects
are ignored and that the opacities as well as the radiative-transfer
codes have obvious shortcomings make it difficult to derive
strong constraints on the explosion mechanism. It appears, however,
that while it seems to be difficult
to distinguish between pure deflagrations and delayed-detonations
on the basis of synthetic lightcurves and spectra, sub-Chandrasekhar
models cannot fit normal SN Ia equally well.
  
\section{PROGENITOR SYSTEMS}
\label{progenitors}
 
In contrast to supernovae from collapsing massive stars for which in two cases
the progenitor star was identified and some of its properties could be inferred 
directly from observations (SN1987A in the LMC \markcite{Bl87,Gi87,GCCe87,HHWe87}({Blanco} 1987; {Gilmozzi} 1987; {Gilmozzi} et~al. 1987; {Hillebrandt} et~al. 1987)
and SN1993J in M81 \markcite{BLHe93,SKEe93,NSSe93,Po93}({Benson} et~al. 1993; {Schmidt} et~al. 1993; {Nomoto} et~al. 1993; {Podsiadlowski} 1993)), there
is not a single case known where we have this kind of information for the progenitor
of a SN Ia. This is not too surprising, given the fact that their
progenitors are most likely faint compact dwarf stars and not red or blue
supergiants. Therefore we have to rely on indirect means if we want to
determine their nature.

The standard procedure is then to eliminate all potential candidates if some
of their properties disagree with either observations or physical
principles, and to hope that one is left with a single and unique
solution. Unfortunately, for the progenitors of Type Ia supernovae
this cannot be done unambiguously, the problem being the {\sl lack} of
strong candidates that pass all possible tests beyond doubt.

In this Section we will first repeat the major constraints which have
to be imposed on the progenitor systems and then discuss the
presently favored candidates, Chandrasekhar-mass C+O white dwarfs and low-mass
C+O white dwarf cores embedded in a shell of helium, in some detail. 
It will be shown, however, that even if we could single out a
particular progenitor system this would narrow the parameter space for
the initial conditions at the onset of the explosion, but might not
determine them sufficiently well, in particular if we are aiming at a
quantitative understanding. Some of the discussion given below
follows recent reviews of \markcite{Re96}{Renzini} (1996) and \markcite{L99}Livio (2000).

\subsection{Observational constraints on Type Ia progenitors}

As was already discussed in Sect. 2, SNe Ia are
(spectroscopically) defined by the absence of emission lines of
hydrogen and the presence of a (blue-shifted) Si II absorption line
with a rest-wavelength of 6355\AA ~near maximum light. The first
finding requires that the atmosphere of the exploding star contains 
no or at most 0.1 M$_{\odot}$ of hydrogen, and the second one indicates
that some nuclear processing takes place and that products of nuclear
burning are ejected in the explosion. Mean velocities of the ejecta,
as inferred from spectral fits, are around 5,000 km/s and peak
velocities exceeding 20,000 km/s are observed, which is consistent
with fusing about 1 M$_{\odot}$ of carbon and oxygen into Fe-group
elements or intermediate-mass elements such as Si or Ca.
The presence of some UV-flux, the width of the peak of the early light
curve, and the fact that radioactive-decay models ($^{56}$Ni $\rightarrow$
$^{56}$Co $\rightarrow$ $^{56}$Fe)  can fit the emission very well, 
all point towards compact
progenitor stars with radii of less than about 10,000 km.

After about two weeks the typical SN Ia spectrum changes from being
dominated by lines of intermediate-mass nuclei to being dominated by Fe
II. Since also a Co III feature is identified at later stages this
adds evidence to the interpretation that they are indeed thermonuclear
explosions of rather compact stars, leaving the cores of stars with 
main sequence masses near 6 to 8 M$_{\odot}$ or white dwarfs as
potential candidates. Moreover, the energetics of the explosion and
the spectra seem to exclude He white dwarfs \markcite{NS77,WTW86}({Nomoto} \& {Sugimoto} 1977; {Woosley}, {Taam}, \& {Weaver} 1986),
mainly because such white dwarfs would undergo very violent
detonations. 
 
Next one notes that most SNe Ia, of order 85 \%,
have very similar peak luminosities, light curves, and spectra. The
dispersion in peak blue and visual brightness is only of order 0.2 -
0.3 magnitudes calling for a very homogeneous class of progenitors. It
is mainly this observational fact that seems to single out
Chandrasekhar-mass white dwarfs as their progenitors. Since
the ratio of energy to mass determines the velocity
profile of the exploding star the homogeneity  would be explained in a
very natural way. However, as has been discussed in Sect. (2),
there exist also significant differences among the various SNe Ia
which may indicate that this simple interpretation is not fully
correct. The difference in peak-brightness, ranging from sub-luminous
events like SN 1991bg in NGC 4374 ($B_{max}$ = -16.54 
\markcite{TBCe96}({Turatto} et~al. 1996), as compared to the mean of the  
``Branch-normals'' of $B_{max} \simeq$ -19  \markcite{HPSe96b}({Hamuy} et~al. 1996c)) 
to bright ones like SN 1991T, which was about 0.5 magnitudes brighter
in $B$ than a typical Type Ia in the Virgo cluster \markcite{MDT95}({Mazzali}, {Danziger}, \& {Turatto} 1995),
is commonly attributed to different $^{56}$Ni-masses
produced in the explosion. They range from about
0.07 M$_{\odot}$ for SN 1991bg ( see, e.g., \markcite{MCTe97}{Mazzali} et~al. (1997a)) to 
at least 0.92 M$_{\odot}$ for SN 1991T (\markcite{KMH93}({Khokhlov} et~al. 1993); see 
however \markcite{FBHe99}{Fisher} et~al. (1999)), with typically 0.6 M$_{\odot}$ for
normal SNe Ia \markcite{HK96,NBBe97}({H\"oflich} \& {Khokhlov} 1996; {Nugent} et~al. 1997).  
It is hard to see how this rather large range can be
accommodated in a single class of models. 

The stellar populations in which SNe Ia show up include
spiral arms as well as elliptical galaxies, with some weak indication
that they might be more efficiently produced in  young populations
\markcite{BTF94}({Bartunov}, {Tsvetkov}, \&  {Filimonova} 1994). Again, if we insist on
a single class of progenitors, the very fact that they do occur in
ellipticals would rule out massive stars as potential candidates. On
the other hand side, the observations may tell us that there is not a
unique class of progenitors. In particular, the fact that the bright
and slowly declining ones (like SN 1991T) are absent in elliptical
and S0 galaxies may point towards different progenitor classes 
\markcite{HPSe96b}({Hamuy} et~al. 1996c).  

All in all, the observational findings summarized so far are
consistent with the assumption that Type Ia supernovae are the result
of thermonuclear disruptions of white dwarfs, C+O white dwarfs being 
the favored model. The diversity 
among them must then be attributed to the history and nature of the
white dwarf prior to the explosion and/or to the physics of
thermonuclear burning during the event. It cannot be excluded,
however, that at least some SNe Ia have a different origin, such as
accretion-induced collapse of massive O-Ne-Mg (or O-Ne) white
dwarfs for SN 1991bg--like objects
\markcite{NYPe94,NIYe95,NYSe96,FBHCe99}({Nomoto} et~al. 1994a, 1995, 1996; {Fryer} et~al. 1999). Also it is 
not clear whether or not there is a clear-cut distinction between Type
Ib/c supernovae, defined by the absence of the Si II feature, and the
(faint) SNe Ia. The former are believed to reflect the core-collapse
of a massive star, its hydrogen-rich envelope being pealed off 
due to mass-loss in a binary system. For example, SN1987K started out
as a SN II with H lines in its spectrum, but changed into a SN
Ib/c-like spectrum after 6 months \markcite{Fi88}({Filippenko} 1988), supporting this
interpretation. It should be noted that SN 1991bg-like objects are not
often observed, but that this may well be a selection effect. \markcite{Su96}{Suntzeff} (1996),
for example, argues that up to 40\% of all Type Ia's could
perhaps belong to that sub-group.

\subsection{Pre-supernova evolution of binary stars}

Despite of all these uncertainties it is the current understanding and
believe that the progenitors of SNe Ia are C+O white dwarfs
in binary systems evolving to the stage of explosion by mass-overflow
from the companion (single-degenerate scenario) or by the merger of
two white dwarfs (double-degenerate scenario). Binary evolution of
some sort is necessary because C+O white dwarfs a typically born with
a mass around 0.6 M$_{\odot}$ \markcite{HKHe98}({Homeier} et~al. 1998) but need to be near the
Chandrasekhar mass or to accumulate a shell of helium in order to explode.  
In this Subsection we will summarize the arguments in favor and
against both scenarios.

Double-degenerates as potential Type Ia progenitors had many ups and
downs in the past, beginning with the classic papers of \markcite{IT84}{Iben} \& {Tutukov} (1984)
and \markcite{W84}{Webbink} (1984). The arguments in favor are that
such binaries should exist as a consequence of stellar evolution,
they would explain very naturally the absence of hydrogen, and they
could, in principle, be an easy way to approach a critical mass. In
fact, several candidate-systems of binary white dwarfs have recently been
identified but most of the short-period ones (at present 8 
systems are known with orbital periods of
less than half a day), which could merge in a Hubble-time due to the
emission of gravitational radiation, have a mass less than \mch 
(\markcite{SLY98}{Saffer}, {Livio}, \& {Yungelson} (1998); see also \markcite{L99}Livio (2000) for a recent
review). There is only one system known (KPD 0422+5421; \markcite{KOW98}{Koen}, {Orosz}, \& {Wade} (1998))
with a mass which, within the errors, could exceed
\mch, a surprisingly small number. None-the-less it is argued
that from population synthesis one could arrive at about the right
frequency of sufficiently massive mergers \markcite{L99}(Livio 2000).
 
Besides the lack of convincing direct observational evidence for
sufficiently many appropriate binary systems, the homogeneity of
``typical'' SNe Ia may be an argument against this class of
progenitors. It is not easy to see how the merging of two white dwarfs
of (likely) different mass, composition, and angular momentum
with different impact parameters, etc., will always lead to
the same burning conditions and, therefore, the production of a nearly
equal amount of $^{56}$Ni. Moreover, 
some investigations of white dwarf mergers seem to indicate 
that an off-center ignition will convert carbon and oxygen into
oxygen, neon, and magnesium, leading to gravitational collapse rather
than a thermonuclear disruption \markcite{WW86a,SN85,SN98,ML90}({Woosley} \& {Weaver} 1986a; {Saio} \& {Nomoto} 1985, 1998; {Mochkovitch} \& {Livio} 1990).
Finally, based on their galactic chemical
evolution model, \markcite{KTNe98}{Kobayashi} et~al. (1998) claim that double-degenerate
mergers lead to inconsistencies with the observed O/Fe as a function of
metallicity, but this statement is certainly model-dependent. In any
case, mergers might, if they are not responsible for the bulk of the
SNe Ia, still account for some peculiar ones, such as the
super-luminous SN 1991T -like explosions.
 
Single-degenerate models are in general favored today.
They consist of a low-mass white dwarf accreting matter
from the companion-star until either it reaches 
\mch or a layer of helium has formed on-top of its C+O
core that can ignite and possibly drive a burning front into the
carbon and oxygen fuel. This track to 
thermonuclear explosions of white dwarfs was first discussed by 
\markcite{WI73,N82a,IT84}{Whelan} \& {Iben} (1973); {Nomoto} (1982b); {Iben} \& {Tutukov} (1984) and \markcite{P85}{Paczynski} (1985). The major problem of these
models has always been that nearly all possible accretion rates can
be ruled out by rather strong arguments 
\markcite{MR92,CCT96,TY96,LBYe96,KVT98,KH99,CIT98}({Munari} \& {Renzini} 1992; {Cassisi}, {Castellani}, \& {Tornambe} 1996; {Tutukov} \& {Yungelson} 1996; {Livio} et~al. 1996; {King} \& {Van Teeseling} 1998; {Kato} \& {Hachisu} 1999; {Cassisi}, {Iben}, \& {Tornambe} 1998).
In short, it is believed that
white dwarfs accreting hydrogen at a low rate undergo nova eruptions
and lose more mass in the outburst than they have accreted prior to it
(e.g. \markcite{Be74,GTW93}{Beer} (1974); {Gehrz}, {Truran}, \& {Williams} (1993)). At moderate accretion rates, a degenerate layer of
helium is thought to form which might flash and could give rise to
sub-Chandrasekhar explosions (which have other problems, as will be
discussed later). Next, still higher accretion rates can lead to quiet
hydrostatic burning of H and He, but these systems should be so bright
that they could easily be detected, but it is not clear beyond doubt
that they coincide with any of the known symbiotic or cataclysmic
binaries. Very high accretion rates, finally would form an
extended H-rich red giant envelope around the white dwarf the debris
of which are not seen in the explosions \markcite{NNS79}({Nomoto}, {Nariai}, \& {Sugimoto} 1979) (see also Fig. 2).
Therefore, it is very uncertain if white dwarfs accreting hydrogen
from a companion star can ever reach the \mch \markcite{CIT98}({Cassisi} et~al. 1998).

\begin{figure*}
\epsfysize=0.80\textwidth\epsfbox{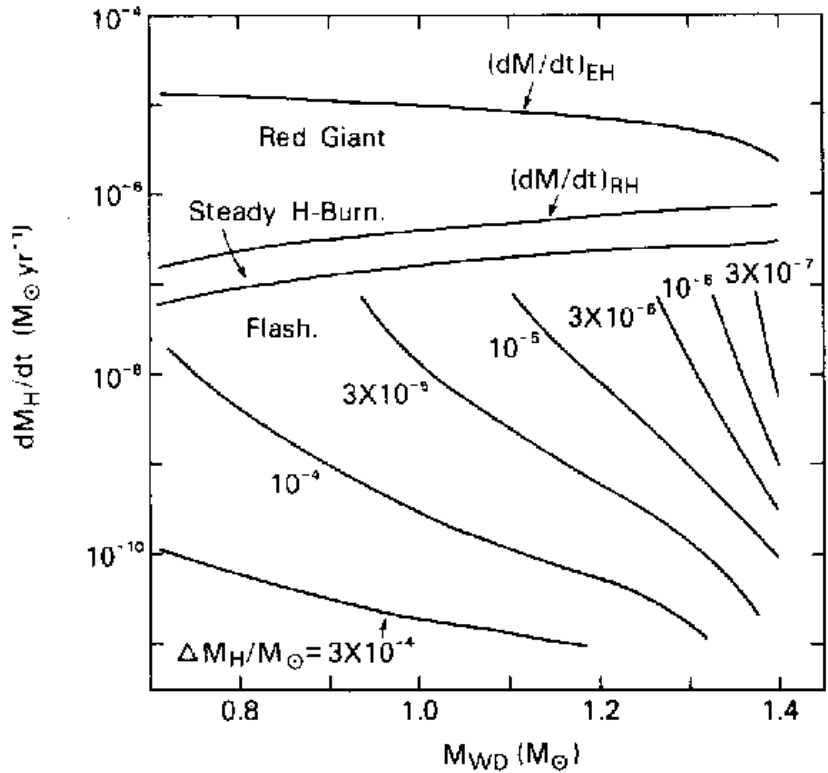}
\caption{\label{fig 4} The likely outcome of hydrogen accretion onto 
white dwarfs of different masses is shown. (From Nomoto 1982a.)}
\end{figure*}

Some of these arguments may be questioned, however. Firstly,
a class of binary systems has recently been discovered, the so-called
``Supersoft X-ray Sources'', which are best interpreted  as white
dwarfs accreting hydrogen-rich matter at such a high rate that H 
burns steadily \markcite{THAe91,GHK91,VDHe92,SLCe96,KH97}({Truemper} et~al. 1991; {Greiner}, {Hasinger}, \& {Kahabka} 1991; {Van Den Heuvel} et~al. 1992; {Southwell} et~al. 1996; {Kahabka} \& {Van Den Heuvel} 1997).
It appears that if these white dwarfs could
retain the accreted gas they might be good candidates for SN Ia progenitors. 
In principle, they could accrete a few tenths of a solar mass with a
typical accretion rate of a few 10$^{-7}$M$_{\odot}$/yr over the estimated 
lifetime of such systems of several 10$^9$ years.
Since most of them are heavily extinct their total
number might be sufficiently high (\markcite{DSR94}{Di Stefano} \& {Rappaport} (1994);
see also \markcite{Li96}{Livio} (1996) and \markcite{YLTe96}{Yungelson} et~al. (1996)),  although this
statement is certainly model-dependent. However, some of the
supersoft X-ray sources are known to be variable in X-rays (but not in
the optical wave-bands) on time-scales of weeks \markcite{PMBe93}({Pakull} et~al. 1993),
too short to be related with the H-burning shell, 
possibly indicating substantial changes in the accretion rates.
It may therefore not be justified to assume that the accretion rates we
see now are sustained over several 10$^9$ years. But their very
existence provides a first and strong case for the single-degenerate
scenario. 

Secondly, also the minimum accretion rate at which hydrogen burns
quietly without a nova outburst is rather uncertain. All models
that compute this rate ignore important pieces of physics and, therefore, 
their predictions could be off by orders of magnitude. For example,
classical nova outbursts require that the accreted hydrogen-rich
envelope of the white dwarf is also heavily enriched in C and O from
the white dwarf's core (see, e.g., \markcite{STSe72,SST76,STS78,Tr82}{Starrfield} et~al. (1972); {Sparks}, {Starrfield}, \& {Truran} (1976); {Starrfield}, {Truran}, \& {Sparks} (1978); {Truran} (1982)).
One possible explanation has been that
convective mixing and dredge-up might happen during the thermonuclear
runaway, but recent numerical simulations indicate that this mechanism
is insufficient \markcite{KHT99}({Kercek}, {Hillebrandt}, \& {Truran} 1999). In contrast to
spherically symmetric models their 3-D simulations lead to a phase of
quiet H-burning for accretion rates as low as 5 $\times$
10$^{-9}$M$_{\odot}$/yr for a white dwarf of 1 M$_{\odot}$ rather than a
nova outburst with mass-loss from the core. Other short-comings
include the assumption of spherical accretion with zero entropy, the
neglect of magnetic fields, etc. So the dividing line between steady
hydrogen burning and nova eruptions might leave some room for SN Ia
progenitors. 

Finally, it has been argued that the interaction of a wind from the
white dwarf with the accretion flow from lobe-filling low mass
red giant may open a wider path to Type Ia supernovae. 
In a series of papers\markcite{HKN96,HKNe99,HKN99}{Hachisu}, {Kato}, \& {Nomoto} (1996); {Hachisu} et~al. (1999b); {Hachisu}, {Kato}, \&  {Nomoto} (1999a) discuss the effect that
when the mass accretion rate exceeds a certain critical value
the envelope solution on the white dwarf is no longer static
but corresponds to a strong wind. The strong wind stabilizes 
the mass transfer and limits the accretion rate and the white dwarf
can burn hydrogen steadily. However, their model assumes spherical
accretion onto and a spherical wind from the white dwarf which seem to
be contradicting assumptions. But the idea should certainly be
followed up. 

\subsection{Evolution to ignition}

In what follows we will assume most of the time that SN Ia progenitors
are Chandrasekhar-mass C+O white dwarfs because,
as was discussed in the previous sections, this class of models
seems to fit best the ``typical'' or ``average'' Type Ia.
In this subsection we also will not discuss models in which 
two degenerate white dwarfs merge and form a critical mass for 
the ignition of carbon, mainly because the merging process will, in 
reality, be very complex and it is difficult to construct 
realistic  explosion models (although with increasing computational
resources it may be possible in the future). 

But even if we consider only Chandrasekhar-mass white dwarfs as
progenitor candidates the information that is needed in order to model
the explosion cannot be obtained easily. In particular, the thermal
structure and the chemical composition are very uncertain.  
The C/O-ratio, for example, has to be know throughout the white dwarf,
but this ratio depends on the main sequence mass of its progenitor and
the metallicity of the gas from which it formed \markcite{UNKe99,WL99}({Umeda} et~al. 1999; {Wellstein} \& {Langer} 1999).
It was found that, depending on the main sequence
mass, the central C/O can vary from 0.4 to 0.6, considerably less than
assumed in most supernova models.

Next, the thermal structure of a white dwarf on its way to an explosion 
depends on the (convective) URCA-process \markcite{Pa73,I78,I82,BW90,M96}({Paczynski} 1973; {Iben} 1978, 1982; {Barkat} \& {Wheeler} 1990; {Mochkovitch} 1996).
The URCA-pairs A = 21, 23, and 25 (such as, i.e., $^{21}$Ne/$^{21}$F,
...) can lead to either heating or cooling, and possibly even to 
a temperature inversion near the center of the white dwarf. The
abundances of the  URCA-pairs depends again on the initial metallicity
which could, thus, affect the thermal structure of the white dwarf.
Unfortunately, the convection in the degenerate star 
is likely to be non-local, time-dependent, 3-dimensional, and very
sub-sonic, but needs to be modeled over very long (secular)
time-scales. It is not likely that in the near future we will be 
able to model these processes in a realistic manner, even on 
super-computers.

Due to these difficulties, numerical studies of the explosion rely
on ad-hoc assumptions fixing the initial conditions, which are usually
chosen to be as simple as possible. Realistic simulations have to be
multi-dimensional, as will be explained in the next Section, and
therefore numerical studies can only investigate a small fraction of
the available parameter space. The failure or success of a particular
model to explain certain observational results may, therefore, not be
conclusive.

\section{EXPLOSION MODELING}

Numerical models are needed to provide the density, temperature,
composition, and velocity fields of the supernova ejecta that result
from the thermonuclear explosion of a white dwarf, accepted by most
researchers as  the ``standard model'' for SNe Ia
(Sec.~(\ref{observations}), (\ref{progenitors})). This information can
then be used to compute the resulting light curve and spectra with the
help of radiation transport codes (Sec.~(\ref{radtrans})) or to compare
the relative distribution of isotopes with the observed solar
abundances.

To a very good approximation, the exploding white dwarf material can
be  described as a fully ionized plasma with varying degrees of
electron degeneracy, satisfying the fluid approximation.  The
governing equations are the hydrodynamical equations for mass,
species, momentum, and energy transport including gravitational
acceleration, viscosity, heat and mass diffusion \markcite{LL95}(Landau \& Lifshitz 1963), and
nuclear energy generation \markcite{A96}(Arnett 1996). They must  be supplemented by
an equation of state for an ideal gas of nuclei, an arbitrarily
relativistic and degenerate electron gas, radiation, and
electron-positron pair production and annihilation \markcite{CG68}(Cox \& Giuli 1968). The
gravitational potential is calculated with the help of the Poisson
equation. In numerical simulations that fully resolve the relevant
length scales  for dissipation, diffusion, and nuclear burning it is
possible to obtain the energy generation rate from a nuclear reaction
network \markcite{T99}(Timmes 1999, for a recent overview, see) and the diffusion
coefficients from an evaluation 
of the kinetic transport mechanisms \markcite{NP84}({Nandkumar} \& {Pethick} 1984). If, on the other
hand, these scales are unresolved -- as  is usually the case in
simulations on scales of the stellar radius -- subgrid-scale models
are required to compute (or parameterize) the effective large-scale
transport coefficients and burning rates, which are more or less
unrelated to the respective microphysical quantities \markcite{K95,NH95a}({Khokhlov} 1995; {Niemeyer} \& {Hillebrandt} 1995b).

Initial conditions can be obtained from hydrostatic spherically
symmetric models of the accreting white dwarf or -- for Chandrasekhar
mass progenitors -- from the Chandrasekhar equation for a fully
degenerate, zero temperature white dwarf \markcite{KW89}(Kippenhahn \& Weigert 1989).  Given the
initial conditions and symmetries specifying the boundary conditions,
the dynamics of the explosion can in principle be determined by
numerically integrating the equations of motion. \markcite{M98}{M{\"u}ller} (1998) gives a
detailed account of some current numerical techniques used for modeling
supernovae.

Until the mid-nineties, most work on SN Ia explosions was done studying
one-dimensional (1D), spherically symmetric models. This approach
inherently lacks some important aspects of multidimensional
thermonuclear burning relevant for \mch-explosion models ,
e.g. off-center flame ignition,  flame instabilities, and turbulence,
which have to be mimicked by means of a spherical flame front with an
undetermined turbulent flame speed
\markcite{NSN76,NTY84,WW86a,W90}e.g., {Nomoto}, {Sugimoto}, \& {Neo} (1976); {Nomoto}, {Thielemann}, \& {Yokoi} (1984); {Woosley} \& {Weaver} (1986a); Woosley (1990). In spite 
of these caveats, 1D models still represent the only reasonable
approach to combine the hydrodynamics with detailed nucleosynthesis
calculations and to carry out parameter studies of explosion
scenarios. In fact, most of the phenomenology of SN Ia explosions and
virtually all of the model predictions for spectra and light curves
are based on spherically symmetric models. Several recent articles
\markcite{W90,NYSe96,HK96,IBNe99}(Woosley 1990; {Nomoto} et~al. 1996; {H\"oflich} \& {Khokhlov} 1996; {Iwamoto} et~al. 1999) describe the methodology and trends
observed in these studies, as well as their implications regarding the
cosmological supernova surveys \markcite{HWT98,RC98,UNKe99,SBB99}({H\"oflich} et~al. 1998b; {Ruiz-Lapuente} \& {Canal} 1998; {Umeda} et~al. 1999; Sorokina, Blinnikov, \& Bartunov 1999).

Following the pioneering work of \markcite{MA82,MA86}{M{\"u}ller} \& {Arnett} (1982, 1986), some groups have
explored the dynamics of two-dimensional (2D)
\markcite{L93,AL94a,AL94b,NH95a,NHW96,A97,RHN99}({Livne} 1993; {Arnett} \& {Livne} 1994a, 1994b; {Niemeyer} \& {Hillebrandt} 1995b; {Niemeyer}, {Hillebrandt}, \&  {Woosley} 1996; {Arnett} 1997; {Reinecke}, {Hillebrandt}, \&  {Niemeyer} 1999a) and three-dimensional
(3D) \markcite{K94,K95,BG97,B97}({Khokhlov} 1994, 1995; {Bravo} \& {Garcia-Senz} 1997; {Benz} 1997) explosion models, triggering the
development of numerical algorithms for representing thin propagating
surfaces in large-scale simulations
\markcite{K93a,NH95a,BG95,A97,GBS98,RHNe99}({Khokhlov} 1993a; {Niemeyer} \& {Hillebrandt} 1995b; {Bravo} \& {Garcia-Senz} 1995; {Arnett} 1997; {Garcia-Senz}, {Bravo}, \&  {Serichol} 1998; {Reinecke} et~al. 1999b). It has also become  possible
to perform 2D and 3D direct numerical simulations (DNS), i.e. fully
resolving the relevant burning and diffusion scales, of microscopic
flame instabilities and flame-turbulence interactions
\markcite{NH95b,K95,NH97,NBR99}({Niemeyer} \& {Hillebrandt} 1995a; {Khokhlov} 1995; {Niemeyer} \& {Hillebrandt} 1997; {Niemeyer}, Bushe, \& Ruetsch 1999).

\subsection{Chandrasekhar Mass Explosion Models}

Given the overall homogeneity of SNe Ia (Sec.~(\ref{properties})), the
good agreement of parameterized 1D \mch-models with observed spectra
and light curves, and their reasonable nucleosynthetic yields, the
bulk of normal SNe Ia is generally assumed to consist of exploding
\mch C+O white dwarfs \markcite{HF60,A69,HW69}({Hoyle} \& {Fowler} 1960; Arnett 1969; Hansen \& Wheeler 1969). In spite of three decades
of work on the hydrodynamics of this  explosion mechanism (beginning
with \markcite{A69}Arnett (1969)), no clear consensus has been reached whether the
star explodes as a result of a subsonic nuclear deflagration that
becomes strongly turbulent \markcite{IIC74,BM75,NSN76,NTY84,WAW84}({Ivanova}, {Imshennik}, \&  {Chechetkin} 1974; {Buchler} \& {Mazurek} 1975; {Nomoto} et~al. 1976, 1984; {Woosley}, {Axelrod}, \& {Weaver} 1984), or
whether this turbulent flame phase is followed by a delayed detonation
during the expansion \markcite{K91a,K91b,WW94a}({Khokhlov} 1991a, 1991b; {Woosley} \& {Weaver} 1994a) or after one or many
pulses \markcite{K91b,AL94a,AL94b}({Khokhlov} 1991b; {Arnett} \& {Livne} 1994a, 1994b). Only the prompt detonation mechanism
is agreed to be inconsistent with SN Ia spectra as it fails to produce
sufficient amounts of intermediate mass elements \markcite{A69,ATW71}(Arnett 1969; {Arnett}, {Truran}, \& {Woosley} 1971).

This apparently slow progress is essentially a consequence of the
overwhelming complexity of turbulent flame physics  and
deflagration-detonation transitions (DDTs) \markcite{W85,ZBLe85}(Williams 1985; Zeldovich et~al. 1985) that
makes first-principle predictions based on \mch-explosion models
nearly impossible. The existence of an initial subsonic flame phase
is, it seems, an unavoidable ingredient of all \mch-models (and only
those) where it is required to pre-expand the stellar material prior to
its nuclear consumption in order to avoid the almost exclusive
production of iron-peaked nuclei \markcite{NSN76,NTY84,WW86a}({Nomoto} et~al. 1976, 1984; {Woosley} \& {Weaver} 1986a).

Guided by parameterized 1D models that yield estimates for the values
for the turbulent flame speed $S_{\rm t}$ and the DDT transition
density $\rho_{\rm DDT}$ \markcite{HK96}(e.g., {H\"oflich} \& {Khokhlov} 1996), a lot of work has been
done recently on the physics of buoyancy-driven, turbulent
thermonuclear flames in exploding \mch-white dwarfs. The close analogy
with thin chemical premixed flames has been exploited to develop a
conceptual framework that covers all scales from the white dwarf
radius to the microscopic flame thickness and dissipation scales
\markcite{K95,NW97}({Khokhlov} 1995; {Niemeyer} \& {Woosley} 1997). In the following discussion of nuclear combustion
(Sec.~(\ref{flame})), flame ignition (Sec.~(\ref{ignite})), and the
various scenarios for \mch explosions characterized by the sequence of
combustion modes (Sec.~(\ref{prompt}) -- Sec.~(\ref{pulsational})) we
will emphasize the current understanding of physical processes rather
than empirical fits of light curves and spectra.

\subsubsection{FLAMES, TURBULENCE, AND DETONATIONS}
\label{flame}

Owing to the strong temperature dependence of the nuclear reaction
rates, $\dot S \sim T^{12}$ at $T \approx 10^{10}$ K \markcite{HK94}(Hansen \& Kawaler 1994, p.~247),
nuclear burning during the 
explosion is  confined to microscopically thin layers that propagate
either conductively as subsonic deflagrations (``flames'') or by shock
compression as supersonic detonations
\markcite{CF48,LL95}(Courant \& Friedrichs 1948; Landau \& Lifshitz 1963, chap.~XIV). Both modes are hydrodynamically unstable
to spatial perturbations as can be shown by linear perturbation
analysis.  In the nonlinear regime, the burning fronts are either
stabilized by forming a cellular structure or become fully turbulent
-- either way, the total burning rate increases as a result of flame
surface growth \markcite{LE61,W85,ZBLe85}(Lewis \& von Elbe 1961; Williams 1985; Zeldovich et~al. 1985). Neither flames nor detonations
can be resolved in explosion simulations on stellar scales and
therefore have to be represented by numerical models.

When the fuel exceeds a critical temperature $T_{\rm c}$ where burning
proceeds nearly instantaneously compared with the fluid motions (see
\markcite{TW92}{Timmes} \& {Woosley} (1992) for a suitable definition of $T_{\rm c}$), a thin
reaction zone forms at the interface between burned and unburned
material. It propagates into the surrounding fuel by one of two
mechanisms allowed by the Rankine-Hugoniot jump conditions: a
deflagration (``flame'') or a detonation (cf.~fig.~2.5 in \markcite{W85}Williams (1985)).

If the overpressure created by the heat of the burning products is
sufficiently high, a hydrodynamical shock wave forms that ignites the
fuel by compressional heating. A self-sustaining combustion front that
propagates by shock-heating is called a detonation. Detonations
generally move supersonically and   therefore do not allow the
unburned medium to expand before it is burned. Their speed depends
mainly on the total amount of  energy released per unit mass,
$\epsilon$, and is therefore more robustly computable than
deflagration velocities. A good estimate for the velocity of planar
strong detonations is the Chapman-Jouget velocity \markcite{LE61,ZBLe85,W85}(Lewis \& von Elbe 1961; Zeldovich et~al. 1985; Williams 1985, and
references therein). The nucleosynthesis, speed,
structure, and stability of  planar detonations in degenerate C+O
material was analyzed by \markcite{IK84,K88,K89,K93b}{Imshennik} \& {Khokhlov} (1984); {Khokhlov} (1988, 1989, 1993b), and recently by
\markcite{KBL98}{Kriminski}, {Bychkov}, \&  {Liberman} (1998) and \markcite{IKKe99}{Imshennik} et~al. (1999) who claim that C+O detonations are
one-dimensionally unstable and therefore cannot occur in exploding
white dwarfs above a critical density of $\sim 2 \times 10^7$ g
cm$^{-3}$ \markcite{KBL98}({Kriminski} et~al. 1998) (cf.~Sec.~(\ref{prompt})).

If, on the other hand, the initial overpressure is too weak, the
temperature gradient at the fuel-ashes interface steepens until an
equilibrium between heat diffusion (carried out predominantly by
electron-ion collisions) and energy generation is reached.  The
resulting combustion front consists of a diffusion zone that heats up
the fuel to $T_{\rm c}$, followed by a thin reaction layer where the
fuel is consumed and energy is generated. It is called a deflagration
or simply a flame and moves subsonically with respect to the unburned
material \markcite{LL95}(Landau \& Lifshitz 1963).  Flames, unlike detonations, may therefore be
strongly affected by turbulent velocity fluctuations of the fuel. Only
if the unburned material is at rest, a unique laminar flame speed
$S_{\rm l}$ can be found which depends on the detailed interaction of
burning and diffusion within the flame region
\markcite{ZBLe85}(e.g., Zeldovich et~al. 1985). Following \markcite{LL95}Landau \& Lifshitz (1963), it can be
estimated by assuming that in order for burning and diffusion to be in
equilibrium, the respective time scales, $\tau_{\rm b} \sim
\epsilon/\dot w$ and $\tau_{\rm d} \sim \delta^2/\kappa$, where
$\delta$ is the flame thickness and $\kappa$ is the thermal
diffusivity, must be similar: $\tau_{\rm b} \sim \tau_{\rm
d}$. Defining $S_{\rm l} = \delta/\tau_{\rm b}$, one finds $S_{\rm l}
\sim (\kappa \dot w/\epsilon)^{1/2}$, where $\dot w$ should be
evaluated at $T \approx T_{\rm c}$ \markcite{TW92}({Timmes} \& {Woosley} 1992).  This is only a crude
estimate due to the strong $T$-dependence of $\dot w$. Numerical
solutions of the full equations of hydrodynamics including nuclear
energy generation and heat diffusion are needed to obtain more
accurate values for $S_{\rm l}$ as a function of $\rho$ and fuel
composition.  Laminar thermonuclear carbon and oxygen flames at high
to intermediate densities were investigated by
\markcite{BCM80,IIC82,WW86b}{Buchler}, {Colgate}, \& {Mazurek} (1980); {Ivanova}, {Imshennik}, \&  {Chechetkin} (1982); {Woosley} \& {Weaver} (1986b), and, using a variety of different
techniques and nuclear networks, by \markcite{TW92}{Timmes} \& {Woosley} (1992). For the purpose of
SN Ia explosion modeling, one needs to know the laminar flame speed
$S_{\rm l} \approx 10^7 \dots 10^4$ cm s$^{-1}$ for $\rho \approx 10^9
\dots 10^7$ g cm$^{-3}$, the flame thickness $\delta = 10^{-4} \dots
1$ cm (defined here as the width of the thermal pre-heating layer
ahead of the much thinner reaction front), and the density contrast
between burned and unburned material $\mu = \Delta \rho/\rho = 0.2
\dots 0.5$ (all values quoted here assume a composition of $X_{\rm C}
= X_{\rm O} = 0.5$, \markcite{TW92}{Timmes} \& {Woosley} (1992)). The thermal expansion  parameter
$\mu$ reflects the partial lifting of electron degeneracy in the
burning products, and is much lower than the typical value found in
chemical, ideal gas systems \markcite{W85}(Williams 1985).

Observed on scales much larger than $\delta$, the internal
reaction-diffusion  structure can be neglected and the flame can be
approximated as a density jump that propagates locally with the normal
speed $S_{\rm l}$. This ``thin flame'' approximation allows a linear
stability analysis of the front with respect to spatial
perturbations. The result shows that thin flames are linearly unstable
on all wavelengths. It was discovered first by \markcite{L44}Landau (1944) and
\markcite{D38}Darrieus (1944) and is hence called the ``Landau-Darrieus'' (LD)
instability.  Subject to the LD instability, perturbations grow until
a web of cellular  structures forms and stabilizes the front at
finite perturbation amplitudes \markcite{Z66}(Zeldovich 1966). The LD instability
therefore does not, in   general, lead to the production of
turbulence. In the context of SN Ia models, the nonlinear LD
instability was studied by \markcite{BS96}Blinnikov \& Sasorov (1996), using a statistical approach
based on the Frankel equation, and by \markcite{NH95b}{Niemeyer} \& {Hillebrandt} (1995a) employing 2D
hydrodynamics and a one-step burning rate. Both groups concluded that
the cellular stabilization mechanism precludes a strong acceleration
of the burning front as a result of the LD instability. However,
\markcite{BS96}Blinnikov \& Sasorov (1996) mention the possible breakdown of stabilization at low
stellar densities (i.e., high $\mu$) which is also indicated by the
lowest density run of \markcite{NH95b}{Niemeyer} \& {Hillebrandt} (1995a) -- this may be important in the
framework of active turbulent combustion (see below). The linear
growth rate of LD unstable thermonuclear flames with arbitrary
equation of state was derived by \markcite{BL95a}{Bychkov} \& {Liberman} (1995a). The same authors also
found a one-dimensional, pulsational instability of degenerate C+O
flames \markcite{BL95b}({Bychkov} \& {Liberman} 1995b) which was later disputed by \markcite{B96}Blinnikov (1996).

The best studied and probably most important hydrodynamical effect for
modeling SN Ia explosions is the Rayleigh-Taylor (RT) instability
\markcite{R83,C61}(Rayleigh 1883; Chandrasekhar 1961) resulting from the buoyancy of hot, burned fluid with
respect to the dense, unburned material. Several groups have
investigated the RT instability of nuclear flames in SNe Ia by means
of numerical hydrodynamical simulations
\markcite{MA82,MA86,L93,K94,K95,NH95a}({M{\"u}ller} \& {Arnett} 1982, 1986; {Livne} 1993; {Khokhlov} 1994, 1995; {Niemeyer} \& {Hillebrandt} 1995b). After more than five decades of
experimental and numerical work, the basic phenomenology of nonlinear
RT mixing is fairly well understood \markcite{F51,L55,S84,R84,Y84}(Fermi 1951; {Layzer} 1955; Sharp 1984; Read 1984; Youngs 1984):
Subject to the RT instability, small surface perturbations grow until
they form bubbles (or ``mushrooms'') that begin to float upward while
spikes of dense fluid fall down. In the nonlinear regime, bubbles of
various sizes interact and create a foamy RT mixing layer whose
vertical extent $h_{\rm RT}$ grows with time $t$ according to a
self-similar growth law, $h_{\rm RT} = \alpha g (\mu/2) t^2$, where
$\alpha$  is a dimensionless constant ($\alpha \approx 0.05$) and $g$
is the background gravitational acceleration \markcite{S84,Y84,R84}(Sharp 1984; Youngs 1984; Read 1984). 
%This growth relation was used by \markcite{NW97}{Niemeyer} \& {Woosley} (1997) to parameterize the
%turbulent flame speed in 1D SN Ia models under the assumption that
%flame propagation is dominated completely by RT mixing
%(cf.~Sec.~(\ref{turbulent})).

Secondary instabilities related to  the velocity shear along the
bubble surfaces \markcite{NH97}({Niemeyer} \& {Hillebrandt} 1997) quickly lead to the production of  turbulent velocity
fluctuations that cascade from the size of the largest bubbles
($\approx 10^7$ cm) down to the microscopic Kolmogorov scale, $l_{\rm
k} \approx 10^{-4}$ cm where they are dissipated
\markcite{NH95a,K95}({Niemeyer} \& {Hillebrandt} 1995b; {Khokhlov} 1995). Since no computer is capable of resolving this
range of scales, one has to resort to statistical or scaling
approximations of those length scales that are not properly
resolved. The most prominent scaling relation in turbulence 
research is Kolmogorov's law for the cascade of velocity fluctuations,
stating that in the case of isotropy and statistical stationarity, the
mean velocity $v$ of turbulent eddies with size $l$ scales as $v \sim
l^{1/3}$ \markcite{K41}(Kolmogorov 1941). 
%Given the velocity of large eddies, e.g. from
%computer simulations, one can use this relation to extrapolate the
%eddy velocity distribution down to smaller scales under the assumption
%of isotropic, fully developed turbulence \markcite{NH95a}({Niemeyer} \& {Hillebrandt} 1995b).
Knowledge of the eddy velocity as a function of length scale is
important to classify the  burning regime of the turbulent combustion
front \markcite{NW97,NK97,KOW97}({Niemeyer} \& {Woosley} 1997; {Niemeyer} \& {Kerstein} 1997; {Khokhlov}, {Oran}, \& {Wheeler} 1997). The ratio of the laminar flame speed
and the turbulent velocity on the scale of the flame thickness, $K =
S_{\rm l}/v(\delta)$, plays an important role: if  $K \gg 1$, the
laminar flame structure is nearly unaffected by turbulent
fluctuations. Turbulence does, however, wrinkle and deform the flame
on  scales $l$ where $S_{\rm l} \ll v(l)$, i.e. above the  {\em Gibson
scale} $l_{\rm g}$ defined by $S_{\rm l} = v(l_{\rm g})$
\markcite{P88}(Peters 1988). These  wrinkles increase the flame surface area and
therefore the total energy generation rate of the turbulent front
\markcite{D40}(Damk{\"o}hler 1940). In other words, the turbulent  flame speed, $S_{\rm t}$,
defined as the mean overall propagation velocity of the turbulent
flame front, becomes larger than the laminar speed $S_{\rm l}$. If the
turbulence is sufficiently strong,  $v(L) \gg  S_{\rm l}$, the
turbulent flame speed becomes independent of the laminar speed, and
therefore of the microphysics of burning and diffusion, and scales
only with the velocity of the largest turbulent eddy \markcite{D40,C94}(Damk{\"o}hler 1940; Clavin 1994):
\begin{equation}
\label{st}
S_{\rm t} \sim v(L)\,\,.
\end{equation}
Because of the unperturbed laminar flame properties on very small
scales, and the wrinkling of the flame on large scales, the burning
regime where $K \gg 1$ is called the corrugated flamelet regime
\markcite{P90,C94}(Pope 1987; Clavin 1994). 
%The properties of the flamelet regime,  in particular
%Eq.~(\ref{st}), form the basis of the model assumptions of the
%multidimensional SN Ia simulations by \markcite{NH95a,NHW96}{Niemeyer} \& {Hillebrandt} (1995b); {Niemeyer} et~al. (1996) and
%\markcite{RHN99}{Reinecke} et~al. (1999a).

As the density of the white dwarf material declines and the laminar
flamelets become slower and thicker, it is plausible that at some
point turbulence significantly alters the thermal flame structure
\markcite{KOW97,NW97}({Khokhlov} et~al. 1997; {Niemeyer} \& {Woosley} 1997). This marks the end of the flamelet regime and
the beginning of the distributed burning, or distributed reaction
zone, regime \markcite{P90}(e.g., Pope 1987). So far, modeling the distributed burning
regime in exploding white dwarfs has not been attempted explicitely
since neither nuclear burning and diffusion nor turbulent mixing can
be properly described by simplified prescriptions. Phenomenologically,
the laminar flame structure is believed to be disrupted by turbulence
and to form a distribution of reaction zones with various lengths and
thicknesses. In order to find the critical density for the transition
between both regimes, we need to formulate a specific criterion for
flamelet breakdown.  A criterion for the transition between both
regimes is discussed in \markcite{NW97,NK97}{Niemeyer} \& {Woosley} (1997); {Niemeyer} \& {Kerstein} (1997) and \markcite{KOW97}{Khokhlov} et~al. (1997):
\begin{equation}
l_{\rm cutoff} \le \delta\,\,.
\end{equation} 
Inserting the results of \markcite{TW92}{Timmes} \& {Woosley} (1992) for $S_{\rm l}$ and $\delta$ as
functions of density, and using a typical turbulence velocity
$v(10^6\mbox{cm}) \sim 10^7$ cm s$^{-1}$, the transition from flamelet
to distributed burning can be shown to occur at a density of
$\rho_{\rm dis} \approx 10^7$ g cm$^{-3}$ \markcite{NK97}({Niemeyer} \& {Kerstein} 1997).

The close coincidence of $\rho_{\rm dis}$ and the preferred value for
$\rho_{\rm DDT}$ \markcite{HK96,NYSe96}({H\"oflich} \& {Khokhlov} 1996; {Nomoto} et~al. 1996) inspired some authors
\markcite{NW97,KOW97}({Niemeyer} \& {Woosley} 1997; {Khokhlov} et~al. 1997) to suggest that  both are related by local flame
quenching and re-ignition via the Zeldovich induction time gradient
mechanism \markcite{ZLMe70}(Zeldovich et~al. 1970), whereby a macroscopic region with a uniform
temperature gradient can give birth to a supersonic spontaneous
combustion wave that steepens into a detonation \markcite{W90}(Woosley 1990, and references
therein). In the context of the SN Ia explosion mechanism, this
effect was first analyzed by  \markcite{BK86,BK87}{Blinnikov} \& {Khokhlov} (1986, 1987). Whether or not the
gradient mechanism can account for DDTs in the delayed detonation
scenario for SNe Ia is still controversial; while \markcite{KOW97}{Khokhlov} et~al. (1997)
conclude that it can, \markcite{N99}{Niemeyer} (1999) -- using arguments based on
incompressible computations of microscopic flame-turbulence
interactions by \markcite{NBR99}{Niemeyer} et~al. (1999) -- states that thermonuclear flames may be
too robust with respect to turbulent quenching to allow the formation
of a sufficiently uniform temperature gradient.

Assuming that the nonlinear RT instability dominates the turbulent
flow that advects the flame, the passive-surface description of the
flame neglects  the additional stirring caused by thermal expansion
within the flame brush itself, accelerating the burnt material in
random directions.  Both the spectrum and cutoff scale may be affected
by ``active'' turbulent combustion \markcite{K96,NW97}(Kerstein 1996; {Niemeyer} \& {Woosley} 1997).  Although the
small expansion coefficient $\mu$ indicates that the effect is weak
compared to chemical flames, a quantitative answer is still missing.

Finally, we note that some authors also studied the multidimensional
instability of detonations in degenerate C+O matter
\markcite{BWOe96,GWKe99}({Boisseau} et~al. 1996; {Gamezo} et~al. 1999), finding unsteady front propagation, the
formation of a cellular front structure and locally incomplete burning
in multidimensional C+O detonations. These effects may have
interesting implications for SN Ia scenarios involving a detonation
phase.

\subsubsection{FLAME IGNITION}
\label{ignite}

As the white dwarf grows close to the Chandrasekhar mass \mch $\approx
1.4 M_\odot$,  the energy budget near the core is governed by plasmon
neutrino losses and compressional heating. The neutrino losses
increase with growing central density until the latter reaches
approximately $2 \times 10^9$ g cm$^{-3}$ \markcite{WW86a}({Woosley} \& {Weaver} 1986a). At this point,
plasmon creation becomes strongly suppressed while electron screening
of nuclear reactions enhances the energy generation rate until it
begins to exceed the neutrino losses. This ``smoldering'' of the core
region marks the beginning of the thermonuclear runaway
\markcite{A69,A71,WW86a}(Arnett 1969; {Arnett} 1971; {Woosley} \& {Weaver} 1986a).  During the following $\sim$ 1000  years, the
core experiences internally heated convection with progressively
smaller turnover time scales $\tau_{\rm c}$. Simultaneously,  the
typical time scale for thermonuclear burning, $\tau_{\rm b}$, drops
even faster as a result of the rising core temperature and  the steep
temperature dependence of the nuclear reaction rates.

During this period, the entropy and temperature evolution of the core
is affected by the convective URCA process, a convectively driven
electron capture-beta decay cycle leading to neutrino-antineutrino
losses. It was first described in this context by \markcite{P72}Paczynski (1972) who
argued it would cause net cooling and therefore delay the
runaway. Since then, the convective URCA process was revisited by
several authors \markcite{B73,I82,BW90,M96}(e.g, {Bruenn} 1973; {Iben} 1982; {Barkat} \& {Wheeler} 1990; {Mochkovitch} 1996) who alternately
claimed that it results in overall heating or cooling. The most recent
analysis \markcite{SBW99}({Stein}, {Barkat}, \& {Wheeler} 1999) concludes that while the  URCA neutrinos
carry away energy, they cannot cool the core globally but instead slow
down the convective motions.

At $T \approx 7 \times 10^8$~K, $\tau_{\rm c}$ and $\tau_{\rm b}$
become comparable,  indicating that convective plumes burn at the same
rate as they circulate \markcite{NTY84,WW86a}({Nomoto} et~al. 1984; {Woosley} \& {Weaver} 1986a). Experimental or numerical
data describing this regime of strong reactive convection is not
available, but several groups are planning to conduct numerical
experiments at the time this article is written. At $T \approx 1.5
\times 10^9$ K, $\tau_{\rm b}$ becomes extremely  small compared with
$\tau_{\rm c}$, and carbon and oxygen virtually burn in place.  A new
equilibrium between energy generation and transport is found on much
smaller length scales, $l \approx 10^{-4}$ cm, where thermal
conduction by degenerate electrons balances nuclear energy input
\markcite{TW92}({Timmes} \& {Woosley} 1992). The flame is born.

The evolution of the runaway immediately prior to ignition of the
flame is crucial for determining its initial location and shape. Using
a simple toy model, \markcite{GW95}{Garcia-Senz} \& {Woosley} (1995) found that under certain conditions,
burning bubbles subject to buoyancy and drag forces can rise a few
hundred km before flame formation, suggesting a high probability for
off-center ignition at multiple, unconnected points.  As a
consequence, more material burns at lower densities, thus producing
higher amounts of intermediate mass elements than a centrally ignited
explosion. In a parameter study, \markcite{NHW96}{Niemeyer} et~al. (1996) and \markcite{RHN99}{Reinecke} et~al. (1999a)
demonstrated the significant influence of  the location and number of
initially ignited spots on the final explosion energetics and
nucleosynthesis.

\subsubsection{PROMPT DETONATION}
\label{prompt}

The first hydrodynamical simulation of an exploding \mch-white dwarf
\markcite{A69}(Arnett 1969) assumed that the thermonuclear combustion commences as a
detonation wave, consuming the entire star at the speed of
sound. Given no time to expand prior to being burned, the C+O material
in this scenario is transformed almost completely into iron-peaked
nuclei and thus fails to produce significant amounts of intermediate
mass elements, in contradiction to observations \markcite{F97b,F97a}({Filippenko} 1997a, 1997b). It
is for this reason that prompt detonations are generally considered
ruled out as viable candidates for the SN Ia explosion mechanism.

In addition to the empirical evidence, the ignition of a detonation in
the high density medium of the white dwarf core was argued to be an
unlikely event. In spite of the smallness of the critical mass for
detonation at $\rho \approx 2 \times 10^9$ g cm$^{-3}$
\markcite{NW97,KOW97}({Niemeyer} \& {Woosley} 1997; {Khokhlov} et~al. 1997) and the correspondingly large number of critical
volumes in the core ($\sim 10^{18}$), the stringent uniformity
condition for the temperature gradient of the runaway region
\markcite{BK86,BK87}({Blinnikov} \& {Khokhlov} 1986, 1987) was shown to be violated even by the minute amounts
of heat dissipated by convective motions \markcite{NW97}({Niemeyer} \& {Woosley} 1997). A different
argument against the occurrence of a prompt detonation in C+O white
dwarf cores was given by \markcite{KBL98}{Kriminski} et~al. (1998), who found that C+O detonations
may be subject to self-quenching at high material densities ($\rho > 2
\times 10^7$ g cm$^{-3}$) \markcite{IKKe99}(see also {Imshennik} et~al. 1999).

\subsubsection{PURE TURBULENT DEFLAGRATION}
\label{turbulent}

Once ignited (Sec.~(\ref{ignite})), the subsonic thermonuclear flame
becomes highly convoluted as a result of turbulence produced by the
various flame instabilities (Sec.~(\ref{flame})). It  continues to
burn through the star until it either transitions into a detonation or
is quenched by expansion. The key questions with regard to explosion
modeling are: a) What is the effective turbulent flame speed $S_{\rm
t}$ as a function of time, b) Is the total amount of energy released
during the deflagration phase enough to unbind the star and produce a
healthy explosion, and c) Does the resulting ejecta composition and
velocity agree with observations?

By far the most work has been done on 1D models, ignoring the
multidimensionality of the flame physics and instead parameterizing
$S_{\rm t}$ in order to answer b) and c) above \markcite{WW86a,NYSe96}(see {Woosley} \& {Weaver} 1986a; {Nomoto} et~al. 1996, for
reviews). One of the most successful examples, model W7
of \markcite{NTY84}{Nomoto} et~al. (1984), clearly demonstrates the excellent agreement of
``fast'' deflagration models with SN Ia spectra and light
curves. $S_{\rm t}$ has been parameterized differently by different
authors, for instance as a constant fraction of the local sound speed
\markcite{HK96,IBNe99}({H\"oflich} \& {Khokhlov} 1996; {Iwamoto} et~al. 1999), using time-dependent convection theory
\markcite{NSN76,BM75,NTY84,WAW84}({Nomoto} et~al. 1976; {Buchler} \& {Mazurek} 1975; {Nomoto} et~al. 1984; {Woosley} et~al. 1984),  or with a phenomenological fractal
model describing the multiscale character of the wrinkled flame
surface \markcite{W90,W97}(Woosley 1990; {Woosley} 1997b). All of these studies essentially agree that
very good agreement with the observations is obtained if $S_{\rm t}$
accelerates up to roughly 30 \% of the sound speed. There remains a
problem with the overproduction of neutron rich iron-group isotopes in
fast deflagration models \markcite{WAW84,TNY86,IBNe99}({Woosley} et~al. 1984; {Thielemann}, {Nomoto}, \& {Yokoi} 1986; {Iwamoto} et~al. 1999), but this may be alleviated in
multiple dimensions (see below). Turning this argument around,
\markcite{W97b}{Woosley} (1997a) argues that $^{48}$Ca can only be produced by carbon
burning in the very high density regime of a \mch white dwarf core,
providing a clue that a few SNe Ia need to be \mch
explosions igniting at $\rho \ge 2 \times 10^9$ g cm$^{-3}$. A
slightly different approach to 1D 
SN Ia modeling was taken by \markcite{NW97}{Niemeyer} \& {Woosley} (1997), employing the self-similar
growth rate of RT mixing regions (Sec.~(\ref{flame})) to prescribe the
turbulent flame speed. Here, all the free parameters are fixed by
independent simulations or experiments. The result shows a successful
explosion, albeit short on intermediate mass elements, suggesting that
the employed flame model is still too simplistic.

A number of authors have studied multidimensional deflagrations in
exploding \mch-white dwarfs using a variety of hydrodynamical methods
\markcite{L93,AL94a,K95,NH95a,NHW96,RHN99}({Livne} 1993; {Arnett} \& {Livne} 1994a; {Khokhlov} 1995; {Niemeyer} \& {Hillebrandt} 1995b; {Niemeyer} et~al. 1996; {Reinecke} et~al. 1999a). The problem of simulating
subsonic flames in large-scale simulations has two aspects: the
representation of the thin, propagating surface separating hot and
cold material with different densities, and the prescription of the
local propagation velocity $S_{\rm t}(\Delta)$ of this surface as a
function of the hydrodynamical state of the large-scale calculation
with numerical resolution $\Delta$. The former problem has been
addressed with artificial reaction-diffusion fronts in PPM
\markcite{K95,NH95a,NHW96}({Khokhlov} 1995; {Niemeyer} \& {Hillebrandt} 1995b; {Niemeyer} et~al. 1996) and SPH \markcite{GBS98}({Garcia-Senz} et~al. 1998), a PPM-specific flame
tracking  technique \markcite{A97}({Arnett} 1997), and a hybrid flame capuring/tracking
method based on level sets \markcite{RHNe99}({Reinecke} et~al. 1999b) (see Fig. 3).
Regarding the flame speed
prescription, some authors assigned the local front  propagation
velocity assuming that the flame is laminar on unresolved scales $l <
\Delta$ \markcite{AL94a}({Arnett} \& {Livne} 1994a), by postulating that $S_{\rm t}(\Delta)$ is
dominated by the terminal rise velocity of $\Delta$-sized bubbles
\markcite{K95}({Khokhlov} 1995), or by using Eq.~(\ref{st}) together with a subgrid-scale
model for the unresolved turbulent kinetic energy providing
$v(\Delta)$ \markcite{NH95a,NHW96,RHN99}({Niemeyer} \& {Hillebrandt} 1995b; {Niemeyer} et~al. 1996; {Reinecke} et~al. 1999a).

\begin{figure*}
%\begin{tabular}{cc}
{\epsfxsize=0.50\textheight\epsfbox{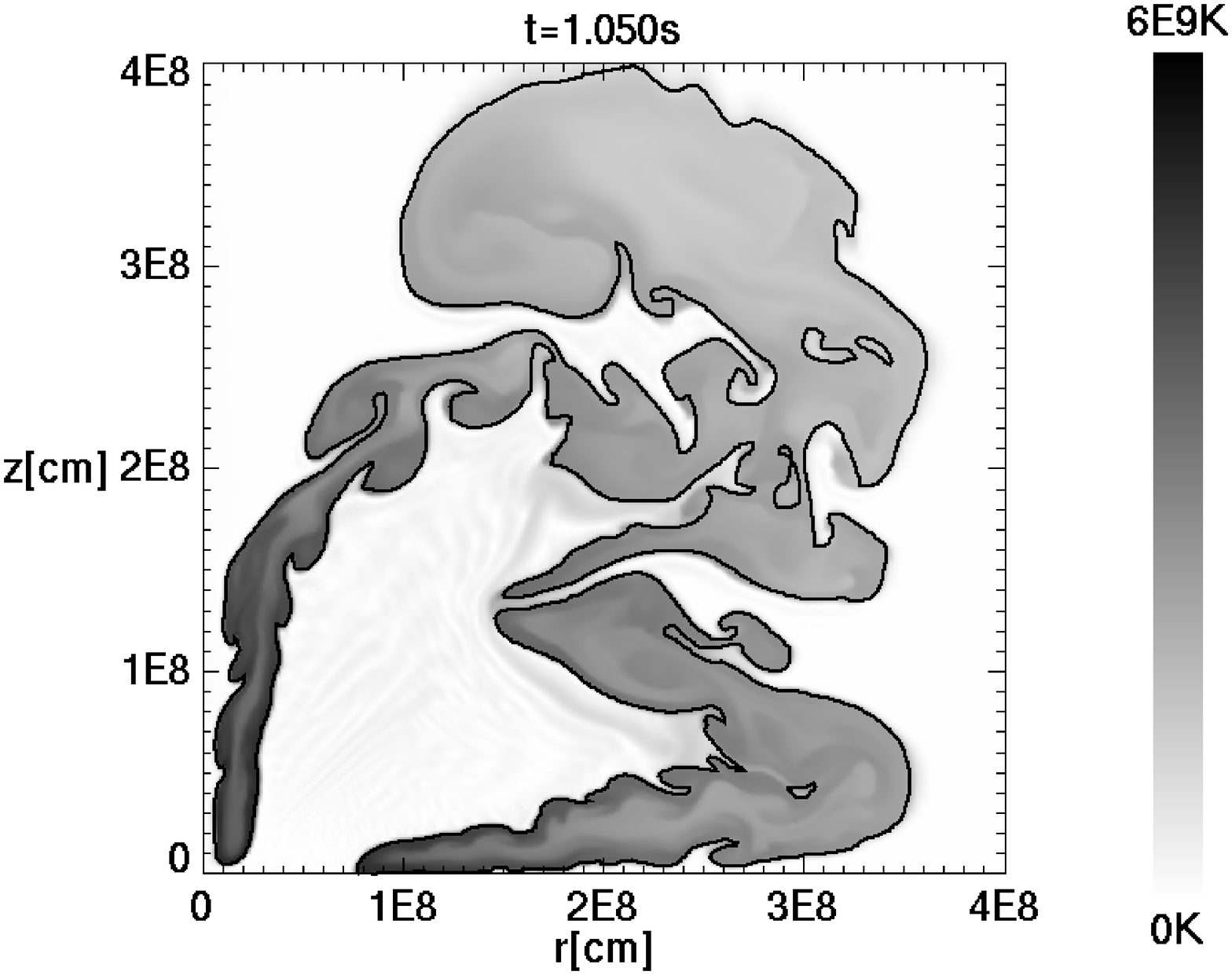}}
%&
{\epsfxsize=0.50\textheight\epsfbox{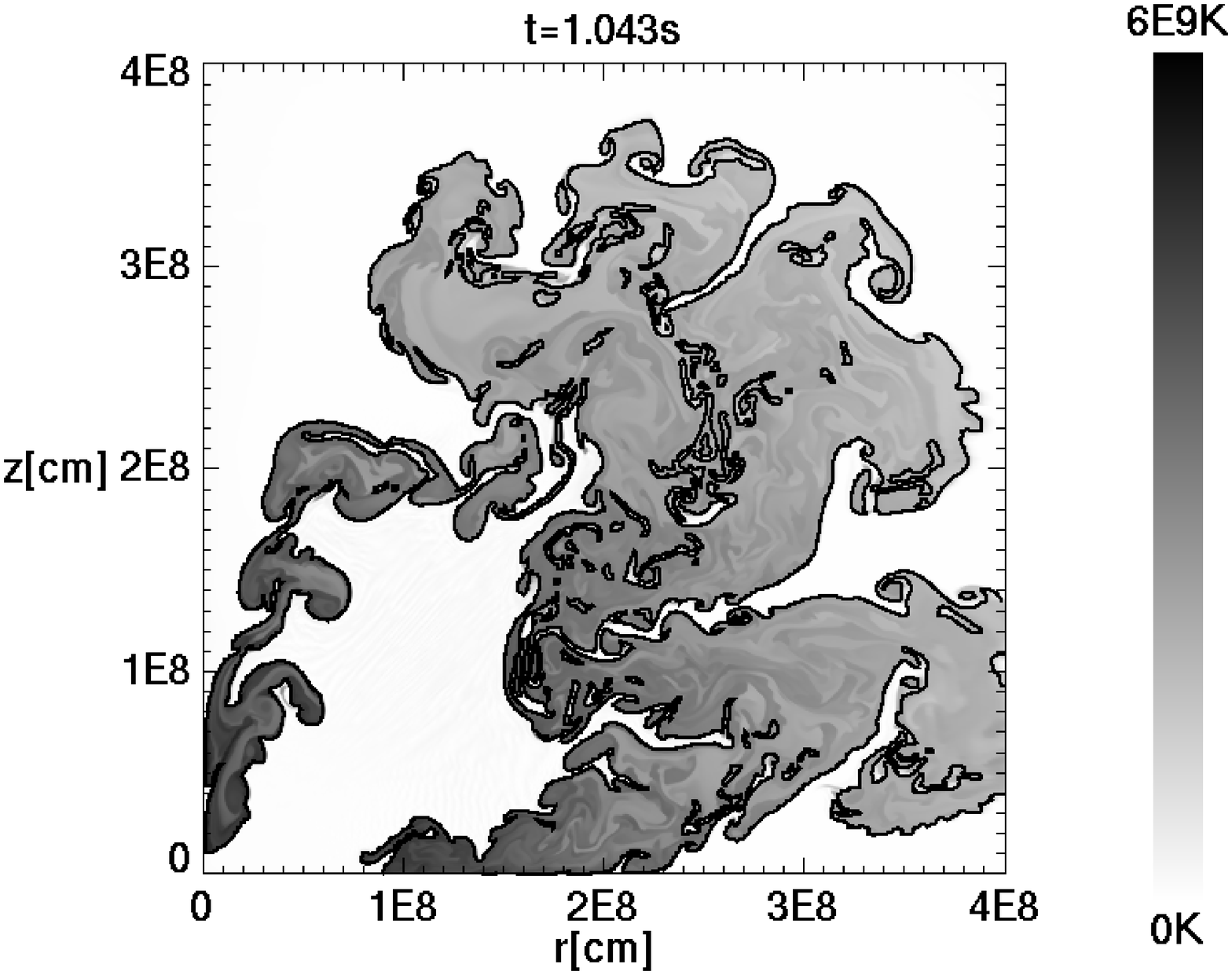}}                           
%\end{tabular}
\caption{Snapshots of the temperature and the front geometry 
in a Chandrasekhar-mass deflagration model at 1.05s (from 
Reinecke et al. (1999c)). Shown are a model with ``low'' resolution
(256$^2$) (upper figure) and one with three times higher resolution,
respectively. Due to the larger surface area of the better resolved
model it exploded, whereas the other one remained marginally bound.}
\label{snapsh}
\end{figure*}
 
In most multidimensional calculations on stellar scales to date, the
effective turbulent flame speed stayed below the required 30 \% of the
sound speed. The detailed outcome of the explosion is controversial;
while some calculations show that the star remains gravitationally
bound after the deflagration phase has ceased \markcite{K95}({Khokhlov} 1995), others
indicate that $S_{\rm t}$ may be large enough to produce a weak but
definitely unbound explosion \markcite{NHW96}({Niemeyer} et~al. 1996). These discrepancies can
probably be attributed to differences in the description of the
turbulent flame and to numerical resolution effects that plague all
multidimensional calculations.
%\markcite{xxx} () report an
%interesting trend with regard to the latter. Tripling the
%grid resolution of an otherwise failing 2D explosion model with
%$256^2$ zones, the total energy release was observed to rise by xxx
%\% as a result of the significant enhancement of turbulent entrainment
%of unburned material in the high-resolution simulation, giving rise to
%a successful explosion. 

\markcite{NW97}{Niemeyer} \& {Woosley} (1997) and \markcite{N99}{Niemeyer} (1999) speculate about additional physics that
can increase the burning rate in turbulent deflagration models, in
particular multipoint ignition and active turbulent combustion (ATC),
i.e.~the generation of additional turbulence by thermal expansion
within the turbulent flame brush. ATC can, in principle, explain the
acceleration of $S_{\rm t}$ up to some fraction of the sound speed
\markcite{K96}(Kerstein 1996),  but its effectiveness is so far unknown. Multipoint
`ignition, on the other hand, has already been shown to significantly
increase the total energy release compared to single-point ignition
models \markcite{NHW96,RHN99}({Niemeyer} et~al. 1996; {Reinecke} et~al. 1999a). Furthermore, it allows more material to
burn at lower densities, thus alleviating the nucleosynthesis problem
of 1D fast deflagration models \markcite{NHW96}({Niemeyer} et~al. 1996).

We conclude the discussion of the pure turbulent deflagration scenario
with a checklist of the model requirements summarized in
Sec.~(\ref{requ}). Assuming that some combination of buoyancy, ATC,
and multipoint ignition can drive the effective turbulent flame speed
to $\sim 30 \%$ of the sound speed -- which is not evident from
multidimensional simulations -- one can conclude from 1D simulations
that pure deflagration models readily comply with all observational
constraints. Most authors agree that $S_{\rm t}$ decouples from
microphysics on large enough scales and becomes dominated by
essentially universal hydrodynamical effects, making the scenario
intrinsically robust. A noteworthy exception is the location and
number of ignition points that can strongly influence the explosion
outcome and may be a possible candidate for the mechanism giving rise
to the explosion strength variability. Other possible sources of
variations include the ignition density and the accretion rate of the
progenitor system \markcite{UNKe99,IBNe99}({Umeda} et~al. 1999; {Iwamoto} et~al. 1999). All of these effects may
potentially vary with composition and metallicity and can therefore
account for the dependence on the progenitor stellar population.

\subsubsection{DELAYED DETONATION}
\label{delayed}

Turbulent deflagrations can sometimes be observed to undergo
spontaneous transitions to detonations (deflagration-detonation
transitions, DDTs) in terrestrial combustion experiments
\markcite{W85}(e.g., Williams 1985, pp.~217--219). Thus inspired, it   was suggested
that DDTs may occur in the late phase of a \mch-explosion, providing
an elegant explanation for the initial slow burning required to
pre-expand the star, followed by a fast combustion mode that produces
large amounts of high-velocity intermediate mass elements
\markcite{K91a,WW94a}({Khokhlov} 1991a; {Woosley} \& {Weaver} 1994a). Many 1D simulations have meanwhile demonstrated
the capability of the delayed detonation scenario to provide excellent
fits to SN Ia spectra and light curves \markcite{W90,HK96}(Woosley 1990; {H\"oflich} \& {Khokhlov} 1996), as well as
reasonable nucleosynthesis products with regard to solar abundances
\markcite{K91b,IBNe99}({Khokhlov} 1991b; {Iwamoto} et~al. 1999). In the best fit models, the initial flame phase
has a  rather slow velocity of roughly one percent of the sound speed
and transitions to detonation at a density of $\rho_{\rm DDT} \approx
10^7$ g cm$^{-3}$ \markcite{HK96,IBNe99}({H\"oflich} \& {Khokhlov} 1996; {Iwamoto} et~al. 1999). The transition density was also
found to be a convenient parameter to explain the observed sequence of
explosion strengths \markcite{HK96}({H\"oflich} \& {Khokhlov} 1996).

Various mechanisms for DDT were discussed in the early literature
on delayed detonations (see \markcite{NW97}{Niemeyer} \& {Woosley} (1997) and references
therein). Recent investigations have focussed on the induction time
gradient mechanism \markcite{ZLMe70,LKY78}(Zeldovich et~al. 1970; Lee, Knystautas, \& Yoshikawa 1978), analyzed in the context of
SNe Ia by \markcite{BK86}{Blinnikov} \& {Khokhlov} (1986) and \markcite{BK87}{Blinnikov} \& {Khokhlov} (1987).  It was realized by
\markcite{KOW97}{Khokhlov} et~al. (1997) and \markcite{NW97}{Niemeyer} \& {Woosley} (1997) that a necessary criterion for this
mechanism is the local disruption of the flame sheet by turbulent
eddies, or, in other words, the transition of the burning regime from
``flamelet'' to ``distributed'' burning (Sec.~(\ref{flame})). Simple
estimates \markcite{NK97}({Niemeyer} \& {Kerstein} 1997) show that this transition should occur at
roughly $10^7$ g cm$^{-3}$, providing a plausible explanation for the
delay of the detonation.

The critical length (or mass) scale over which the temperature
gradient must be held fixed in order to allow the spontaneous
combustion wave to turn into a detonation was computed by
\markcite{KOW97}{Khokhlov} et~al. (1997) and \markcite{NW97}{Niemeyer} \& {Woosley} (1997); it is a few orders of magnitude
thicker than the final  detonation front and depends very sensitively
on composition and density. 
%However, \markcite{N99}{Niemeyer} (1999) argues  that it is
%unlikely that turbulence can quench a thermonuclear flame brush over
%such large distances, owing to the robustness of these flames
%\markcite{NBR99}({Niemeyer} et~al. 1999). In addition, it was shown by \markcite{MKO98} () that even
%small perturbations ($\sim 10 \%$) of the gradient region inhibit the
%formation of the detonation. At the present time, one can neither
%exclude nor confirm  the occurrence of DDTs in exploding white dwarfs,
%but they may be as (un)natural as the acceleration of a turbulent
%flame to a fraction of the sound speed.

The virtues of the delayed detonation scenario can again be summarized
by completing the checklist of Sec.~(\ref{requ}). It is undisputed
that suitably tuned delayed detonations satisfy all the constraints
given by SN Ia spectra, light curves, and nucleosynthesis. If
$\rho_{\rm DDT}$ is indeed determined by the transition of burning
regimes  -- which in turn might be composition dependent
\markcite{UNKe99}({Umeda} et~al. 1999) --
the scenario is also fairly robust and $\rho_{\rm DDT}$ may represent
the explosion strength parameter. Note that in this case, the
variability induced by multipoint ignition needs to be explained
away. If, on the other hand, thermonuclear flames are  confirmed to be
almost unquenchable, the favorite mechanism for DDTs becomes
questionable \markcite{N99}({Niemeyer} 1999). Moreover, should the mechanism DDT rely on
rare, strong turbulent fluctuations one  must ask about those events
that fail to ignite a detonation following the slow deflagration phase
which, on its own, cannot give rise to a viable SN Ia explosion. They
might end up as pulsational delayed detonations or as unobservably
dim, as yet unclassified explosions. Multidimensional simulations of
the turbulent flame phase  may soon answer whether the turbulent flame
speed is closer to 1 \% or 30 \% of the speed of sound and hence
decide whether DDTs are a necessary ingredient of SN Ia explosion
models.

\subsubsection{PULSATIONAL DELAYED DETONATION}
\label{pulsational}

In this variety of the delayed detonation scenario, the first
turbulent deflagration phase fails to release enough energy to unbind
the star which subsequently pulses and triggers a detonation upon
recollapse \markcite{NSN76,K91b}({Nomoto} et~al. 1976; {Khokhlov} 1991b). This model was studied in 1D by \markcite{HK96}{H\"oflich} \& {Khokhlov} (1996)
and \markcite{W97}{Woosley} (1997b) (who calls it ``pulsed detonation of the first type'')
and in 2D by \markcite{AL94b}{Arnett} \& {Livne} (1994b). \markcite{HK96}{H\"oflich} \& {Khokhlov} (1996) report that it produces
little $^{56}$Ni but a substantial amount of Si and Ca and may
therefore explain very subluminous events like SN 1991bg.
\markcite{W97}{Woosley} (1997b), using a fractal flame parameterization, also considered
``pulsed deflagrations'', i.e.~re-ignition occurs as a deflagration
rather than a detonation, and ``pulsed detonations of the second
type'' in which the burning also re-ignites as a flame but later
accelerates and touches off a detonation . This latter model closely
resembles the standard delayed detonation, whereas the former may or
may not produce a healthy explosion, depending on the prescribed speed
of the rekindled flame \markcite{W97}({Woosley} 1997b).

Obtaining a DDT by means of the gradient mechanism is considerably
more plausible after one or several pulses than during the first
expansion phase \markcite{KOW97}({Khokhlov} et~al. 1997) as the laminar flame thickness becomes
macroscopically large during the expansion, allowing the fuel to be
preheated, and turbulence is significantly enhanced during the
collapse.

The ``checklist'' for pulsational delayed detonations looks similar to
that of simple delayed detonations (see above), with somewhat less
emphasis on the improbability for DDT. Some fine-tuning of the initial
flame speed is needed to obtain a large enough pulse in order to
achieve a sufficient degree of mixing, while avoiding to unbind the
star in a very weak explosion \markcite{NW97}({Niemeyer} \& {Woosley} 1997). Again, these ``fizzles''
may be very subluminous and may have escaped discovery. We finally
note that all pulsational models are in conflict with multidimensional
simulations that predict an unbound star after the first deflagration
phase.

\subsection{Sub-Chandrasekhar Mass Models}
\label{subchandra}

C+O white dwarfs below the Chandrasekhar mass do not reach the
critical density and temperature for explosive carbon burning by
accretion, and therefore need to be ignited by an external
trigger. Detonations in the accreted He layer were suggested to drive
a strong enough shock into the C+O core to initiate a secondary carbon
detonation \markcite{WW80,N80,N82b,WWT80,SW84,IT84}({Weaver} \& {Woosley} 1980; {Nomoto} 1980, 1982a; {Woosley}, {Weaver}, \& {Taam} 1980; {Sutherland} \& {Wheeler} 1984; {Iben} \& {Tutukov} 1984). The nucleosynthesis and
light curves of Sub-\mch models, also known as ``helium  ignitors'' or
``edge-lit detonations'', were investigated in 1D \markcite{WW94b,HK96}({Woosley} \& {Weaver} 1994b; {H\"oflich} \& {Khokhlov} 1996)
and 2D \markcite{LA95}({Livne} \& {Arnett} 1995) and found to be superficially consistent with SNe Ia,
especially subluminous ones \markcite{RJCe93}({Ruiz-Lapuente} et~al. 1993a). Their ejecta structure is
characterized almost inevitably by an outer layer of high-velocity Ni
and He above the intermediate mass elements and the inner Fe/Ni core.

These models are favored mostly by the statistics of possible SN Ia
progenitor systems \markcite{YL98,L99}({Yungelson} \& {Livio} 1998; Livio 2000) and by the straightforward
explanation of the one-parameter strength sequence in terms of the WD
mass \markcite{RBC95}({Ruiz-Lapuente}, {Burkert}, \&  {Canal} 1995). However, they appear to be severely challenged
both photometrically and spectroscopically: Owing to the heating by
radioactive $^{56}$Ni in the outer layer they are somewhat too blue at
maximum brightness and their light curve rises and declines too
steeply \markcite{HK96,NBBe97,HKWe97}({H\"oflich} \& {Khokhlov} 1996; {Nugent} et~al. 1997; {H\"oflich} et~al. 1997). Perhaps even more stringent is
the generic prediction of He-ignitors to exhibit signatures of
high-velocity Ni and He, rather than Si and Ca, in the early and
maximum spectra which is in strong disagreement with observations
\markcite{NBBe97,HKWe97}({Nugent} et~al. 1997; {H\"oflich} et~al. 1997).

With respect to the explosion mechanism itself, the most crucial
question is whether and where the He detonation manages to shock the
C+O core sufficiently to create a carbon detonation. 1D models, by
virtue of their built-in spherical symmetry, robustly (and
unphysically) predict a perfect convergence of the inward propagating
pressure wave and subsequent carbon ignition near the core
\markcite{WW94b}({Woosley} \& {Weaver} 1994b). Some 2D simulations indicate that the C+O detonation is
born off-center but still due to the convergence of the He-driven
shock near the symmetry axis of the calculation \markcite{L90,LG91}({Livne} 1990; {Livne} \& {Glasner} 1991) while
others find a direct initiation of the carbon detonation along the
circle where the He detonation intersects the C+O core
\markcite{L97,A97,WF97,WSF98}({Livne} 1997; {Arnett} 1997; {Wiggins} \& {Falle} 1997; {Wiggins}, {Sharpe}, \& {Falle} 1998).  Using 3D SPH simulations, \markcite{B97}{Benz} (1997)
failed to see carbon ignition in all but the highest resolution
calculations, where carbon was ignited directly at the interface
rather than by shock convergence. Further, C ignition is facilitated
if the He detonation starts at some distance above the interface,
allowing the build-up of a fully developed pressure spike before it
hits the carbon \markcite{B97}({Benz} 1997). This result was confirmed by recent 3D
SPH simulations \markcite{GBW99}({Garcia-Senz}, {Bravo}, \&  {Woosley} 1999) that also examined the effect of
multiple He ignition points, finding enhanced production of
intermediate mass elements in this case. Hence, multidimensional
SPH and PPM simulations presently confirm the validity of He-driven carbon
detonations, in particular by direct ignition, but they also
demonstrate the need for very high numerical resolution in order to
obtain mutually consistent results \markcite{A97,B97}({Arnett} 1997; {Benz} 1997).

To summarize, sub-\mch models are most severly constrained by their
prediction of an outer layer of high-velocity Ni and He. Should
further research conclude that spectra, colors, and light curves are
less contaminated by this layer than presently thought, they represent
an attractive class of candidates for SNe Ia, especially subluminous
ones, from the point of view of progenitor statistics and the
one-parameter explosion strength family. Note, however, that the SN Ia
luminosity function in this scenario is directly linked to the
distribution of white dwarf masses, predicting a more gradual decline
on the bright side of the luminosity function than indicated by
observations \markcite{VBMe95,L99}({Vaughan} et~al. 1995; Livio 2000). The explosion mechanism itself
appears realistic, at least in the direct carbon ignition mode, but
more work is needed to firmly establish the conditions for ignition of
the secondary carbon detonation.

\subsection{Merging White Dwarfs}
\label{merging}

The most obvious strength of the merging white dwarfs, or
double-degenerate, scenario for SNe Ia \markcite{W84,IT84,P85}({Webbink} 1984; {Iben} \& {Tutukov} 1984; {Paczynski} 1985) is the
natural explanation for the lack of hydrogen in SN Ia spectra
\markcite{L99}(Livio 2000) (cf.~Sec.~(\ref{properties})). Furthermore, in contrast to
the elusive  progenitor systems for single degenerate scenarios, there
is meanwhile some evidence for the existence of double degenerate
binary systems \markcite{SLY98}({Saffer} et~al. 1998) despite earlier suspicions to the
contrary \markcite{Br97}(e.g., Bragaglia 1997). These systems are bound to merge as a
consequence of gravitational wave emission with about the right
statistics \markcite{L99}(Livio 2000) and give rise to some extreme astrophysical
event, albeit not necessarily a SN Ia.

Spherically symmetric models of detonating merged systems,
parameterized as C+O white dwarfs with thick envelopes, were analyzed
by \markcite{HKM92,KMH93}{H\"oflich}, {Khokhlov}, \& {M\"uller} (1992); {Khokhlov} et~al. (1993) and \markcite{HK96}{H\"oflich} \& {Khokhlov} (1996), giving reasonable agreement
with SN Ia light curves. 3D SPH simulations of white dwarfs mergers
\markcite{BCPe90,RS95,MGS97}({Benz} et~al. 1990; {Rasio} \& {Shapiro} 1995; Mochkovitch, Guerrero, \& Segretain 1997) show the  disruption of the less massive
star in a matter of a few orbital times, followed by the formation of
a thick hot accretion disk around the more massive companion. The
further evolution hinges crucially on the effective accretion rate of
the disk: In case $\dot M$ is larger than a few times $10^{-6}$
M$_\odot$ yr$^{-1}$, the most likely outcome is off-center carbon
ignition leading to an inward propagating flame that converts the star
into O+Ne+Mg \markcite{NI85,SN85,KSN87,TWT94,SN98}({Nomoto} \& {Iben} 1985; {Saio} \& {Nomoto} 1985; {Kawai}, {Saio}, \& {Nomoto} 1987; {Timmes}, {Woosley}, \& {Taam} 1994; {Saio} \& {Nomoto} 1998). This configuration,
in  turn, is gravitationally unstable owing to electron capture onto
$^{24}$Mg and will undergo accretion-induced collapse (AIC) to form a
neutron star \markcite{SN85,ML90,NK91}({Saio} \& {Nomoto} 1985; {Mochkovitch} \& {Livio} 1990; {Nomoto} \& {Kondo} 1991). A recent re-examination of
Coulomb corrections to the equation of state of material in nuclear
statistical equilibrium indicates that AIC in merged white dwarf
systems is even more likely than previously anticipated \markcite{BG99}({Bravo} \& {Garcia-Senz} 1999).

Dimensional analysis of the expected turbulent viscosity due to MHD
instabilities \markcite{BHS96}({Balbus}, {Hawley}, \& {Stone} 1996) suggests that it is very difficult to
avoid such high accretion rates \markcite{ML90,L99}({Mochkovitch} \& {Livio} 1990; Livio 2000). Even under the
unphysical assumption that angular momentum transport is dominated
entirely by microscopic electron-gas viscosity, the expected life time
of $\sim 10^9$ yrs \markcite{ML90,MGS97}({Mochkovitch} \& {Livio} 1990; Mochkovitch et~al. 1997) and high UV luminosity of these
accretion systems would predict the existence of $\sim 10^7$ such
objects in the Galaxy, none of which have been observed \markcite{L99}(Livio 2000).

A possible solution to the collapse problem is to ignite carbon
burning as a detonation rather than a flame immediately during the
merger event, either in the core of the more massive star
\markcite{SNYe92}({Shigeyama} et~al. 1992) or at the contact surface (D Arnett \& PA Pinto,
private communication). This alternative clearly warrants further study.

To summarize, the merging white dwarf scenario has to overcome the
crucial problem of avoiding accretion-induced collapse before it can
be seriously considered as a SN Ia candidate. Its key strengths are a
plausible explanation for the progenitor history yielding reasonable
predictions for SN Ia rates, the straightforward explanation of the
absence of H and He in SN Ia spectra, and the existence of a simple
parameter for the explosion strength family (i.e., the mass of the
merged system).

\section{SUMMARY}

In this review we have outlined our present understanding
of Type Ia supernovae, summarizing briefly the observational
constraints, but putting more weight on models of the explosion.
From the tremendous amount of work carried out over the last couple
of years it has become obvious that the physics of SNe Ia is very
complex, ranging from the possibility of very different
progenitors to the complexity of the physics leading to the
explosion and the complicated processes which couple the interior
physics to observable quantities. None of these problems is fully
understood yet, but what one is tempted to state is that, from a 
theorist's point of view, it appears to be a miracle that all 
the complexity seems to average out in a mysterious way 
to make the class so homogeneous. In contrast, as it stands,
a save prediction from theory seems to be that SNe Ia should get
more divers with increasing observed sample sizes. 
If, however, homogeneity would continue to hold
this would certainly add support to the Chandrasekhar-mass 
single-degenerate scenario. On the other hand, even an increasing
diversity would not rule out Chandrasekhar-mass
single-degenerate progenitors for most of them. In contrast,
there are ways to explain how the diversity  
is absorbed in a one parameter family of transformations, such as
the Phillips-relation or modifications of it. For example,  
we have argued that the size of the convective core of 
the white dwarf prior to the explosion might provide a physical
reason for such a relation.

As far as the explosion/combustion physics and the numerical
simulations are concerned significant recent progress has made 
the models more realistic (and reliable). Thanks to ever increasing
computer resources 3-dimensional simulations have become feasible
which treat the full star with good spatial resolution and realistic
input physics. Already the results of 2-dimensional simulations
indicate that pure deflagrations waves in Chandrasekhar-mass C+O white 
dwarfs can lead to explosions, and one can expect that going to
three dimensions, because of the increasing surface area of the
nuclear flames, should add to the explosion energy. If confirmed,
this would eliminate pulsational detonations from the list of
potential models. On the side of the combustion physics, the 
burning in the distributed regime at low densities needs to be
explored further, but it is not clear anymore whether a transition
from a deflagration to a detonation in that regime is needed for
successful models. In fact, according to recent studies such a transition 
appears to be rather unlikely.

Finally, sub-Chandrasekhar-mass models seem to face problems, both
from the observations and from theory, leaving us with the conclusion
that we seem to be lucky and Nature was kind to us and singled 
out from all possibilities the simplest solution, namely a
Chandrasekhar-mass C+O white dwarf and a nuclear deflagration wave,
to make a Type Ia supernova explosion. 

%\bibliography{}

\end{document}